\documentclass[aps,pra,twocolumn,superscriptaddress,nofootinbib,floatfix]{revtex4}

\usepackage{amsmath,amssymb,bm}
\usepackage{graphicx}
\usepackage{booktabs}
\usepackage{dcolumn}
\usepackage{hyperref}
\hypersetup{hidelinks}

\usepackage{CJK}
\usepackage{epsfig}
\usepackage{subfigure}

\newcommand{\ee}{\mathrm{e}}

\newcommand{\Fref}[1] {Figure \ref{#1}}

\newcommand{\vs}{\vspace*}
\newcommand{\ms}{\vspace*{-5mm}}

\begin{document}

\title{Wigner time delay in the $4d$ giant-resonance region of iodoalkanes}

\author{Anatoli S. Kheifets}
\email{A.Kheifets@anu.edu.au}
\affiliation{Research School of Physics, The Australian National University, Canberra ACT 0200, Australia}

\author{Daniele Toffoli}
\affiliation{Dipartimento di Scienze Chimiche e Farmaceutiche, Universita degli Studi di Trieste, Trieste, Italy}

\author{Piero Decleva}
\affiliation{Dipartimento di Scienze Chimiche e Farmaceutiche, Universita degli Studi di Trieste, Trieste, Italy}
\affiliation{CNR-IOM DEMOCRITOS, Trieste, Italy}

\date{\today}

\begin{abstract}
The iodine $4d$ giant resonance provides an unusually strong and
predominantly iodine-centered photoionization channel in iodoalkanes.
Recent attosecond measurements by Schr\"oder et al.  [Phys. Rev. A
  112, L051104 (2025)] reported pronounced, nonmonotonic, and
isomer-dependent variations of the streaking-derived $4d$
photoemission delay across the molecular series.  We calculate
field-free molecular Eisenbud--Wigner--Smith (EWS) delays from
photoionization amplitudes represented in a multicenter B-spline basis
of the time-dependent density-functional theory (TDDFT).  A tensor
formulation is developed in which the delay for randomly oriented
molecules is evaluated from the same rank-0 and rank-2 contractions
that determine the cross section and anisotropy parameter.  The
resulting TDDFT delays of CH$_3$I, C$_2$H$_5$I, 2-iodopropane, and
both calculated conformers of 1-iodopropane form a narrow bundle
throughout the 90--118~eV region and closely follow atomic-I and I$^-$
reference calculations.  This robustness is consistent with the
similarly weak molecular dependence of the giant resonance cross
section and angular distribution, but it is in marked contrast with
the much larger molecular variation found by streaking experiments.
The asymptotic Coulomb--laser-coupling/continuum--continuum correction is
controlled mainly by the photoelectron energy and residual charge and is
therefore expected to be nearly common to the iodoalkane series.  
This does not exclude target-specific molecular IR effects beyond the
universal Coulomb term.
\end{abstract}

\maketitle

\section{Introduction}
\label{sec:intro}

Laser-assisted attosecond photoemission chronoscopy converts the
spectral phase of a photoelectron wave packet into a measurable
temporal quantity. In the absence of the assisting laser field, the
relevant group delay is the Eisenbud--Wigner--Smith (EWS), hereafter
referred to simply as the Wigner delay, defined as the energy
derivative of the complete photoionization phase
\cite{Wigner1955,Pazourek2015,Kheifets2023review}.  Streaking and
RABBITT measurements add an infrared (IR) field to interrogate this
phase \cite{Itatani2002,Yakovlev2010,Dahlstrom2012}.  For simple
atomic targets the probe contribution can often be separated into a
universal Coulomb--laser-coupling (CLC), or continuum--continuum (CC),
term, thereby providing quantitative access to the one-photon delay
\cite{Pazourek2015,Dahlstrom2012}.

Iodine and iodine-containing molecules are attractive targets for such
measurements because of the broad and strong $4d\rightarrow\epsilon f$
giant resonance.  The $d\rightarrow f$ channel dominates the photoelectron
signal over the competing $d\rightarrow p$ channel through the spectral
range around $90$--$120$~eV, while the iodine $4d$ orbitals remain strongly
localized and essentially nonbonding.  Atomic photoionization measurements
and calculations have established the characteristic broad resonance
\cite{Nahon1991,OSullivan1996,PiLandsman2018}, and related giant-resonance
time-delay physics has been studied extensively in xenon
\cite{Magrakvelidze2016,PhysRevA.96.053407,PhysRevA.98.043427}.

This favorable combination of a large cross section and a localized
core orbital motivated the use of iodoalkanes as timing references in
the absolute attosecond streaking experiment of \citet{Ossiander2018}.
\citet{Biswas2020} subsequently interpreted an enhanced I $4d$ delay
in iodoethane as a signature of the molecular environment and proposed
that the photoemission delay should grow with the size of the ligand.
A recent systematic streaking study by \citet{Schroder2025} tested
this proposition in iodomethane, iodoethane, 1- and 2-iodopropane, and
1- and 2-iodobutane at photon energies of 90, 105, and 118~eV.  The
resulting pattern is much richer than a monotonic molecular-size
effect.  In particular, the two propyl isomers, although chemically
identical in composition and similar in their iodine-centered
one-photon ionization, display markedly different streaking delays at
90 and 105~eV.

\begin{figure}[t]
\centering
\includegraphics[width=\columnwidth]{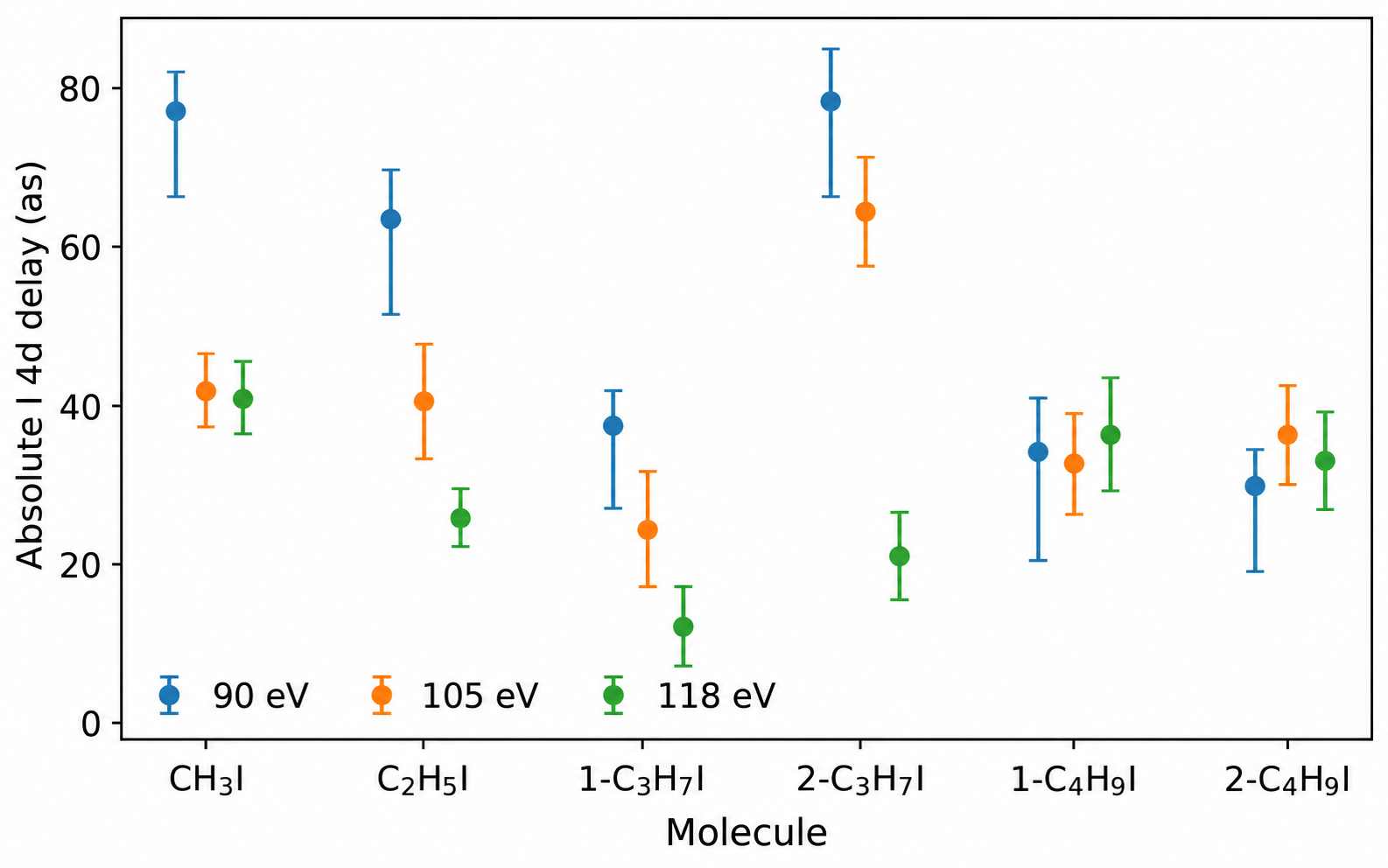}
\caption{Experimental streaking-derived I $4d$ photoemission delays reported
for the iodoalkane series at 90, 105, and 118~eV \cite{Schroder2025}.
The horizontal coordinate labels the molecule; small horizontal offsets separate the three
photon energies.  The pronounced nonmonotonic and isomer-dependent pattern
is the central experimental motivation for the present work.}
\label{fig:experiment}
\end{figure}

The experimental systematics are summarized in
\Fref{fig:experiment}.  The strongest molecular variation occurs where
one would have expected the broad iodine $4d$ giant resonance to be least
sensitive to the ligand.  The effect is especially striking at 90~eV, where
the reported streaking-derived delays range from about 29 as in
2-iodobutane to about 79 as in 2-iodopropane.  At 105~eV the
difference between 1- and 2-iodopropane is about 41 as.  No single ordering of the molecules persists
at all three photon energies \cite{Schroder2025}.

Very recently, \citet{Yao2026} proposed a field-free
single-scattering interpretation of this isomer dependence.  In their
model, the iodine-centered $4d\to f$ photoelectron wave acts as a
reference wave and interferes with components elastically scattered
from the carbon framework.  The resulting structure-dependent Wigner
delays retain a sizeable difference between 1- and 2-iodopropane after
orientational averaging, in qualitative agreement with the measured
isomer contrast.  This result makes the central issue especially sharp:
is the molecular dependence already generated by structural scattering
in the one-photon XUV step, or does the streaking-derived delay also
reflect dynamics beyond the field-free Wigner delay?

A field-free theoretical description of the molecular sequence is
therefore essential for two reasons.  First, it tests whether the XUV
photoionization step itself contains a comparably strong molecular
dependence, including such structural scattering.  Second, it provides the
reference required to identify additional physics introduced by the streaking field.
Previous scattering calculations at the HF+LDA level were found not to
reproduce the giant resonance cross section even for CH$_3$I
\cite{Schroder2025}.  By contrast, the B-spline TDDFT approach used
here has been developed specifically for molecular photoionization
with a correct long-range LB94 potential and has successfully
described molecular-frame photoelectron distributions
\cite{Toffoli2002,Fukuzawa2019}.  Its complex amplitudes contain
multicenter molecular scattering together with linear-response
screening and interchannel coupling, providing a complementary
field-free test of the proposed structural-scattering mechanism.  

In the present work, we first formulate and implement the EWS delay
directly from the complex TDDFT molecular photoionization amplitudes
and identify a previously hidden asymptotic phase convention that is
immaterial for conventional observables but crucial for time delays.
We then compare the corrected molecular delays across the iodoalkane
series with atomic I and I$^-$ benchmarks and with the streaking
measurements.  This comparison directly tests whether the field-free
molecular amplitudes encode the large structure dependence suggested
by the recent single-scattering picture.  The central result is
unexpectedly simple: the field-free molecular delays cluster tightly
around the atomic values, and the giant resonance is remarkably
resilient against chemical substitution.  The large experimental
spread is therefore not reproduced by the present field-free
calculation and points either to physics absent from that description
or to the mapping between the XUV+IR spectrogram and the inferred
one-photon delay. Nor do we recover the sizeable field-free isomer
dependence predicted by the single-scattering model of
\citet{Yao2026}.

An important qualification concerns the IR probe itself.  In a molecule,
subtracting an atomic-like asymptotic CLC/CC term need not remove all
probe-induced phase.  Recent molecular calculations identify anisotropic
continuum--continuum coupling and IR-induced ionic transitions as possible
additional contributions \cite{Delgado2025,Pranjal2026,BendaMasin2024,Makos2025}.
Their applicability and important limitations for the present randomly
oriented, higher-energy streaking experiment are discussed in
Sec.~\ref{sec:molecular-ir}.

Beyond molecular IR-probe effects, the discrepancy also motivates
examination of post-ionization processes absent from the present
IR-field-free calculation.  NVV Auger--Meitner decay is one possibility
because it involves the molecular valence shell, whereas the initial I $4d$
ionization is strongly localized.  However, chemical sensitivity of
the decay does not establish that it produces the measured
primary-photoelectron delay differences.  \citet{Schroder2025}
explicitly tested an overlapping streaked Auger contribution and
included the resulting 6.33 as maximum bias as a systematic
uncertainty at 90~eV \cite{Schroder2025}.  

Post-collision interaction (PCI) is physically distinct from simply
adding two spectra: it couples the outgoing electrons and residual ion
through their Coulomb interactions \cite{Kuchiev1985,Kuchiev1989}.
Its possible importance must nevertheless be assessed using emission
times, velocities, and angular correlations.  The near equality of
primary and Auger electron energies at 90~eV is not sufficient
evidence for enhancement, and the faster primary electrons at 105 and
118~eV make conventional overtaking less favorable.  Field-assisted
PCI theory also distinguishes distortions of the Auger spectrum from
changes of the primary-photoelectron streaking response
\cite{BauchBonitz2012}.  We therefore discuss molecular IR probe
dynamics and Auger/PCI as unconfirmed possibilities without assigning
a hierarchy among them.  The established result of this work is the
weak molecular dependence of the corrected, IR-field-free TDDFT delay,
while the possible origins of the experimentally observed molecular
sensitivity remain speculative.

\section{Theory and computational method}
\label{sec:theory}
\ms

\newcommand{\bi}{\begin{itemize}}
\newcommand{\ei}{\end{itemize}}
\newcommand{\be}{\begin{enumerate}}
\renewcommand{\ee}{\end{enumerate}}
\newcommand{\bt}{\begin{tabular}}
\newcommand{\et}{\end{tabular}}
\newcommand{\bc}{\begin{center}}
\newcommand{\ec}{\end{center}}
\newcommand{\bs}{\begin{slide}{}}
\newcommand{\es}{\end{slide}}
\newcommand{\beq}{\begin{equation}}
\newcommand{\eeq}{\end{equation}}
\newcommand{\td}[1]{ {\rm{\bf #1}} }
\newcommand{\tderiv}[1]{ {\frac{\partial{#1}}{\partial t} }}
\newcommand{\R}[1]{ {\rm #1} }
\newcommand{\B}[1]{ {\bf #1} }
\newcommand{\pop}{ {\td P}_{op}( {\td r}) }
\newcommand{\pin}{{\td P}_{in}({\td r})}
\newcommand{\pope}{{\td P}_{op}^{eq}({\td r})}
\newcommand{\pine}{{\td P}_{in}^{eq}({\td r})}
\newcommand{\ten}{\nabla'\left({1\over {\!\mid\!{\td r}-{\td r}'\!\mid\!}}\right)}
\newcommand{\bea}{\begin{eqnarray}}
\newcommand{\eea}{\end{eqnarray}}
\newcommand{\sss}{\scriptscriptstyle}
\newcommand{\dsty}{\displaystyle}
\newcommand{\spc}{\mbox{\ \ \ \ \ \ \ \ \ \ \ } }
\newcommand{\BB}[1]{\mbox{\boldmath $#1$}}
\newcommand{\expt}[1]{ \langle \tilde{0} | #1 |\tilde{0} \rangle}
\newcommand{\overl}[2]{ \langle #1 |#2 \rangle}
\newcommand{\tmx}[3]{ \langle #1 | #2 |#3 \rangle}
\newcommand{\kap}[1]{\mbox{$\, ^{#1}\!\kappa\! $ }}
\newcommand{\s}[1]{\mbox{$ \,^{#1}\!S\! $ }}
\newcommand{\bra}[1]{\langle #1 |}
\newcommand{\ket}[1]{ |#1 \rangle}
\newcommand{\nod}{ \Psi }
\newcommand{\nodb}{\bar{\nod}}
\newcommand{\nodt}{\tilde{\nod}}
\newcommand{\knod}{ \ket{\nod} }
\newcommand{\knodb}{ \ket{\nodb} }
\newcommand{\knodt}{ \ket{\nodt} }
\newcommand{\bnod}{ \bra{\nod} }
\newcommand{\bnodb}{ \bra{\nodb} }
\newcommand{\bnodt}{ \bra{\nodt} }
\newcommand{\kdnod}{ \ket{\dot{\nod}} }
\newcommand{\tdh}{ \Phi }
\newcommand{\tdhb}{\bar{\tdh}}
\newcommand{\tdht}{\tilde{\tdh}}
\newcommand{\ktdh}{ \ket{\tdh} }
\newcommand{\ktdhb}{ \ket{\tdhb} }
\newcommand{\ktdht}{ \ket{\tdht} }
\newcommand{\btdh}{ \bra{\tdh} }
\newcommand{\btdhb}{ \bra{\tdhb} }
\newcommand{\btdht}{ \bra{\tdht} }
\newcommand{\kdtdh}{ \ket{\dot{\tdh}} }
\newcommand{\tdhmm}{ \tdh^{-m} }
\newcommand{\tdhmmt}{ \tdht^{-m} }
\newcommand{\ktdhmm}{ \ket{\tdh^{-m}} }
\newcommand{\ktdhmmt}{ \ket{\tdht^{-m}} }
\newcommand{\tdhmfh}{ \langle{\oper{H}}\rangle^{-m}}
\newcommand{\vac}{ |{\rm vac} \rangle}
\newcommand{\bvac}{ \langle {\rm vac} |}
\newcommand{\smave}[1]{ \langle #1 \rangle}
\newcommand{\avg}[1]{ \langle #1 \rangle}
\newcommand{\avgO}[1]{ \langle O | #1 |O \rangle}
\newcommand{\tbra}[1]{ \langle O^{#1L} |}
\newcommand{\tket}[1]{  |O^{#1R}  \rangle}
\newcommand{\adj}[1]{  {#1}^{\dagger}}
\newcommand{\phijm}{ {\phi^m_{j^m}(q_m)} }
\newcommand{\lrf}[3]{ \langle\langle #1;#2 \rangle\rangle_{#3}}
\newcommand{\lrfA}[3]{ \langle\langle #1;#2 \rangle\rangle_{#3}^A}
\newcommand{\lrfB}[3]{ \langle\langle #1;#2 \rangle\rangle_{#3}^B}
\newcommand{\lrfst}[2]{ \langle\langle #1;#2 \rangle\rangle_{0}}
\newcommand{\lrfstA}[2]{ \langle\langle #1;#2 \rangle\rangle_{0}^A}
\newcommand{\lrfstB}[2]{ \langle\langle #1;#2 \rangle\rangle_{0}^B}
\newcommand{\lrfstat}[3]{ \langle\langle #1;#2 \rangle\rangle_{0}^{#3}}
\newcommand{\qrf}[5]{ \langle\langle #1;#2,#3 \rangle\rangle_{#4,#5}}
\newcommand{\qrfA}[5]{ \langle\langle #1;#2,#3 \rangle\rangle_{#4,#5}^A}
\newcommand{\qrfB}[5]{ \langle\langle #1;#2,#3 \rangle\rangle_{#4,#5}^B}
\newcommand{\qrfst}[3]{ \langle\langle #1;#2,#3 \rangle\rangle_{0,0}}
\newcommand{\qrfstA}[3]{ \langle\langle #1;#2,#3 \rangle\rangle_{0,0}^A}
\newcommand{\qrfstB}[3]{ \langle\langle #1;#2,#3 \rangle\rangle_{0,0}^B}
\newcommand{\crf}[7]{ \langle\langle #1;#2,#3,#4 \rangle\rangle_{#5,#6,#7}}
\newcommand{\crfA}[7]{ \langle\langle #1;#2,#3,#4 \rangle\rangle_{#5,#6,#7}^A}
\newcommand{\crfB}[7]{ \langle\langle #1;#2,#3,#4 \rangle\rangle_{#5,#6,#7}^B}
\newcommand{\crfst}[4]{ \langle\langle #1;#2,#3,#4 \rangle\rangle_{0,0,0}}
\newcommand{\crfstA}[4]{ \langle\langle #1;#2,#3,#4 \rangle\rangle_{0,0,0}^A}
\newcommand{\crfstB}[4]{ \langle\langle #1;#2,#3,#4 \rangle\rangle_{0,0,0}^B}
\newcommand{\ovev}[1]{ {\bf{#1}} }
\newcommand{\ovevt}[1]{ {{\bf{#1}}^T} }
\newcommand{\cvec}[1]{ {\bf #1} }
\newcommand{\cvecp}[1]{ {\bf #1}^{\prime} }
\newcommand{\cvecpp}[1]{ {\bf #1}^{\prime \prime} }
\newcommand{\cvect}[2]{ ({\bf #1}_{#2})^T}
\newcommand{\cvecd}[1]{ {\dot {\bf #1}}}
\newcommand{\cvecdt}[2]{ ({\dot {\bf #1}}_{#2})^T}
\newcommand{\cvecdp}[1]{ {\dot {\bf #1}}^\prime}
\newcommand{\cvecdpt}[2]{ ({\dot {\bf #1}}^\prime_{#2})^T}
\newcommand{\cvecdpp}[1]{ {\dot {\bf #1}}^{\prime \prime}}
\newcommand{\cvecdppt}[2]{ ({\dot {\bf #1}}^{\prime \prime}_{#2})^T}
\newcommand{\ovem}[1]{ {\bf {#1}}  }
\newcommand{\greekm}[1]{{\BB{#1}} }
\newcommand{\ovemt}[1]{ {\bf {#1}}^T  }
\newcommand{\veclam} {\greekm{\lambda}}
\newcommand{\veclamb} {\bar{\greekm{\lambda}}}
\newcommand{\ovemd}[1]{ {\dot \ovem{#1}}  }
\newcommand{\ovemdt}[1]{ {\dot \ovem{#1}}^T  }
\newcommand{\vdot}{ \cdot}
\newcommand{\schr}{Schr{\"o}dinger }
\newcommand{\cs}{ coordinate system }
\newcommand{\wf}{ wave function }
\newcommand{\wfs}{ wave functions }
\newcommand{\tdse}{ time-dependent Schr{\"o}dinger equation }
\newcommand{\tise}{ time-independent Schr{\"o}dinger equation }
\newcommand{\tdvp}{ time-dependent variational principle }
\newcommand{\half}{\frac{1}{2}}
\newcommand{\mhalf}{{-\frac{1}{2}}}
\newcommand{\third}{\frac{1}{3}}
\newcommand{\fourth}{\frac{1}{4}}
\newcommand{\fifth}{\frac{1}{5}}
\newcommand{\sixth}{\frac{1}{6}}
\newcommand{\eigth}{\frac{1}{8}}
\newcommand{\nfacth}{\frac{1}{n!}}
\newcommand{\sums}{\sum_{s=1}^{N_p}}
\newcommand{\sumi}{\sum_{i=1}^{N_e}}
\newcommand{\sumij}{\sum_{i,j=1}^{N_e}}
\newcommand{\sumprime}[2]{ {\sum_{#1}^{#2}}^\prime}
\newcommand{\sumprimeprime}[2]{ {\sum_{#1}^{#2}}^{\prime\prime}}
\newcommand{\sumP}{\sum_{P=1}^{N_n}}
\newcommand{\sumPp}{\sum_{P^\prime=1}^{N_n}}
\newcommand{\sumk}{\sum_{k=1}^{3N_n-6}}
\newcommand{\suml}{\sum_{l=1}^{3N_n-6}}
\newcommand{\sumlp}{\sum_{l^\prime=1}^{3N_n-6}}
\newcommand{\sumkl}{\sum_{k,l=1}^{3N_n-6}}
\newcommand{\sumkllp}{\sum_{k,l,l^\prime=1}^{3N_n-6}}
\newcommand{\summ}{\sum_{m=1}^{M}}
\newcommand{\sumrm}{\sum_{r^m=1}^{N^m}}
\newcommand{\sumrsm}{\sum_{r^m,s^m=1}^{N^m}}
\newcommand{\sumrlsm}{\sum_{r^m>s^m=1}^{N^m}}
\newcommand{\sumsm}{\sum_{s^m=1}^{N^m}}
\newcommand{\sumK}{\sum_{K=1}^{3N}}
\newcommand{\deltaf}[1]{   \delta_{#1}}
\newcommand{\alphap} {{\alpha^\prime}}
\newcommand{\alphapp}{{\alpha^{\prime\prime}}}
\newcommand{\betap}  {{\beta^\prime}}
\newcommand{\betapp} {{\beta^{\prime\prime}}}
\newcommand{\gammap} {{\gamma^\prime}}
\newcommand{\gammapp}{{\gamma^{\prime\prime}}}
\newcommand{\deltap} {{\delta^\prime}}
\newcommand{\chip}   {{\chi^\prime}}
\newcommand{\xip}    {{\xi^\prime}}
\newcommand{\zetap}  {{\zeta^\prime}}
\newcommand{\etap}   {{\eta^\prime}}
\newcommand{\epsilonp}   {{\epsilon^\prime}}
\newcommand{\preup}[2] {\,^{#1}\!\!{#2}}
\newcommand{\binc}[2] {{\left( \begin{array}{c} {#1} \\ {#2} \end{array} \right) }}
\newcommand{\eqtwoopt}[5]
{
\bea
{#1} =
\left\{
\begin{array}{lll}
{#2}&{#3} \\
{#4}&{#5}
\end{array}
\right.
\eea
}
\newcommand{\threej}[6]
{
\left(
\begin{array}{lll}
{#1}&{#2}&{#3} \\
{#4}&{#5}&{#6}
\end{array}
\right)
}
\newcommand{\VL} {\preup{V}{L}}
\newcommand{\TL} {\,^T\!\!L}
\newcommand{\RL} {\,^R\!\!L}
\newcommand{\VP} {\,^V\!\!P}
\newcommand{\RVP} {\,^{RV}\!\!P}
\newcommand{\RJ} {\,^{R}\!\!J}
\newcommand{\RVJ} {\,^{RV}\!\!J}
\newcommand{\kp} {{k^\prime}}
\newcommand{\Pp} {{P^\prime}}
\newcommand{\pp} {{\prime\prime}}
\newcommand{\oper}[1]{ {\hat{#1}} }
\newcommand{\unitvec}{ \ovem{1} }
\newcommand{\kivec}{ \ovem{k} }
\newcommand{\livec}{ \ovem{l} }
\newcommand{\iivec}{ \ovem{i} }
\newcommand{\jivec}{ \ovem{j} }
\newcommand{\sivec}{ \ovem{s} }
\newcommand{\rivec}{ \ovem{r} }
\newcommand{\pivec}{ \ovem{p} }
\newcommand{\civec}{ \ovem{c} }
\newcommand{\divec}{ \ovem{d} }
\newcommand{\prodm}{ {\prod_{m=1}^M} }
\newcommand{\prodmp}{ {\prod_{m^\prime=1}^M} }
\newcommand{\prodmpp}{ {\prod_{m^{\prime\prime}=1}^M} }
\newcommand{\chivm}{ {\chi^m_{v^m}(q_m)} }
\newcommand{\chium}{ {\chi^m_{u^m}(q_m)} }
\newcommand{\phim}{ {\phi^m(q_m)} }
\newcommand{\phimt}{ {\tilde{\phi}^m(q_m)} }
\newcommand{\phimtt}{ {\tilde{\tilde{\phi}}^m(q_m)} }
\newcommand{\phirm}{ {\phi^m_{r^m}(q_m)} }
\newcommand{\phism}{ {\phi^m_{s^m}(q_m)} }
\newcommand{\phipm}{ {\phi^m_{p^m}(q_m)} }
\newcommand{\phiim}{ {\phi^m_{i^m}(q_m)} }
\newcommand{\phiimp}{ {\phi^{m^\prime}_{i^{m^\prime}}(q_{m^\prime})} }
\newcommand{\phiimpp}{ {\phi^{m^{\prime\prime}}_{i^{m^{\prime\prime}}}(q_{m^{\prime\prime}}}) }
\def\Section#1{{\large\bf{\underline{#1\rule[-1mm]{0mm}{0mm}}}}}
\newcommand{\naba}{   \nabla_{\alpha}}
\newcommand{\nabaone}[1]{ \nabla_{\alpha_{#1}}}
\newcommand{\nabatwo}[2]{ \nabla_{\alpha_{#1}}\nabla_{\alpha_{#2}} }
\newcommand{\alphav}[1]{ {\BB{\alpha}_{#1}}    }
\newcommand{\nabav}[1]{   {\nabla_{\alphav{#1}}}   }
\newcommand{\mcwt}[1]{   {\,^{cwt}\!\!M_{\alphav{#1}}}   }
\newcommand{\mcwtA}[1]{   {\,^{cwt}\!\!M^A_{\alphav{#1}}}   }
\newcommand{\mcwtB}[1]{   {\,^{cwt}\!\!M^A_{\alphav{#1}}}   }
\newcommand{\mctl}[1]{   {\,^{ctl}\!\!M_{\alphav{#1}}}   }
\newcommand{\mctlA}[1]{   {\,^{ctl}\!\!M^A_{\alphav{#1}}}   }
\newcommand{\mctlB}[1]{   {\,^{ctl}\!\!M^B_{\alphav{#1}}}   }
\newcommand{\msg}[1]{   {\,^{sg}\!\!M_{{#1}}}   }
\newcommand{\msgA}[1]{   {\,^{sg}\!\!M^A_{{#1}}}   }
\newcommand{\msgB}[1]{   {\,^{sg}\!\!M^B_{{#1}}}   }
\newcommand{\msp}[1]{   {\,^{s}\!\!M_{{#1}}}   }
\newcommand{\mspA}[1]{   {\,^{s}\!\!M^A_{{#1}}}   }
\newcommand{\mspB}[1]{   {\,^{s}\!\!M^B_{{#1}}}   }
\newcommand{\mop}[1]{   {\oper{M}_{#1} }}
\newcommand{\mtot}[1]{   {M}^{tot}_{#1} }
\newcommand{\mtotA}[1]{   {M}^{A,tot}_{#1} }
\newcommand{\mtotB}[1]{   {M}^{B,tot}_{#1} }
\newcommand{\mopA}[1]{   {\oper{M}_{#1}^A }}
\newcommand{\mopB}[1]{   {\oper{M}_{#1}^B }}
\newcommand{\cgt}[2]{   {\,^{cg}\!T^{#1}_{#2}}}
\newcommand{\ces}{   \frac{1}{4\pi\epsilon_0}        }
\newcommand{\RF}[3]   { {\cal RF}({#1};{#2};{#3})  }
\newcommand{\RFat}[4]   { {\cal RF}^{#4}({#1};{#2};{#3})  }
\newcommand{\RFst}[2] { {\cal RF}({#1};{#2};{\ovev{0}})  }
\newcommand{\RFstat}[3] { {\cal RF}^{#3}({#1};{#2};{\ovev{0}})  }


\newcommand{\CHI}{CH$_3$I}
\newcommand{\CEI}{CH$_3$CH$_2$I}
\newcommand{\CPIiso}{CH$_3$CHICH$_3$}
\newcommand{\CPInorm}{CH$_3$CH$_2$CH$_2$I}
\newcommand{\Iminus}{I$^-$}

\newcommand{\ce}[1]{%
  \ifx\relax#1\relax\else
    \csname chem@#1\endcsname
  \fi
}

\expandafter\def\csname chem@CH3I\endcsname{CH$_3$I}
\expandafter\def\csname chem@CH3CH2I\endcsname{CH$_3$CH$_2$I}
\expandafter\def\csname chem@CH3CHICH3\endcsname{CH$_3$CHICH$_3$}
\expandafter\def\csname chem@CH3CH2CH2I\endcsname{CH$_3$CH$_2$CH$_2$I}
\expandafter\def\csname chem@I-\endcsname{I$^-$}






\subsection{Computational details}

All molecular continuum and TDDFT calculations reported here were
performed with the \textsc{Tiresia} program suite
\cite{Decleva2022Tiresia,Toffoli2024Tiresia}.
The underlying computational methodology has been documented in detail
in Refs.~\cite{jcp_1.1937367,jcp_1.2178799}; here we give only a brief
overview of the elements most relevant to the present calculations.  A
preliminary geometry optimization of each system was carried out with
the ADF engine of the Amsterdam Modeling Suite~\cite{ADF2001,AMS2022},
at the DFT level using the hybrid B3LYP
functional~\cite{Becke_1993,Stephens_1994} and a triple-$\xi$ plus
polarization basis set (TZP) of Slater-type orbitals (STO) taken from
the AMS basis set library.  For each system at its computed
equilibrium geometry, the self-consistent-field (SCF) ground state
electron density is obtained by a conventional bound-state calculation
employing a TZP STO basis and the LB94 exchange-correlation (xc)
potential~\cite{PhysRevA.49.2421}. 
For each molecule the full molecular point-group symmetry is employed
throughout the calculation: $C_{3v}$ for CH$_3$I, $C_s$ for CH$_3$CH$_2$I,
CH$_3$CHICH$_3$, and the lowest-energy conformer of
CH$_3$CH$_2$CH$_2$I, while a slightly higher-energy eclipsed
conformation of CH$_3$CH$_2$CH$_2$I has no symmetry elements and the
$O_h$ point group has been used for the anion I$^-$. The density (and
therefore the scattering potential) is expanded in the LCAO basis set,
which is constructed as follows. Primitive basis functions, built as
products of a radial B-spline~\cite{deBoor_book} and a real spherical
harmonic, are placed at a common origin (a long-range single-center
expansion which describes the asymptotic region) and at all nuclear
centers, and are further symmetry adapted by taking proper linear
combinations. In this work, B-splines of order 12 have been employed
and the single-center expansion, always centered on the iodine atom,
has been built over a linear radial grid extending up to 25.0
a.u. with a step size of 0.25 a.u. This value has been increased to 35
a.u. for I$^-$. Additional knots have been put near the positions of
the off-center nuclei. For all molecular systems, the single center expansion has been truncated at a maximum angular momentum value of $l_{\mathrm{max}}=12$ while this value has been reduced to $l_{\mathrm{max}}=3$ for I$^-$. The B-spline basis set on the off-center nuclei
is defined over a linear radial grid extending up to a value of
$R_{\mathrm{max}}$, which has been set in the interval 0.5--0.7 a.u.
for H and 1.2--1.5 a.u. for C. The maximum angular momentum employed
in the off-center spheres was equal to 2 for the C atoms and 1 for the
H atoms.

Bound state solutions are obtained with a generalized diagonalization
of the Kohn-Sham Hamiltonian matrix, whereas scattering states are
extracted by using a least-squares approach~\cite{BROSOLO1992185}.  In
the LR-TDDFT formalism, the linear response of the electronic density
to a weak time-dependent external electromagnetic field with frequency
$\omega$ is evaluated following the original scheme of Zangwill and
Soven~\cite{PhysRevA.21.1561}: an effective self-consistent field
potential $v^{SCF}$ is introduced which consists of the external
dipole potential, $v^{\textit{ext}}$ plus the Coulomb and
exchange-correlation (xc) screening due to the redistribution of the
electrons:

\begin{equation}
v^{\textit{SCF}}(\mathbf{r},\omega)=v^{\textit{ext}}(\mathbf{r},\omega)+\int\frac{\delta n(\mathbf{r}^{\prime},\omega)}{|\mathbf{r}-\mathbf{r}^{\prime}|}d\mathbf{r}^{\prime}+\frac{\partial v_{xc}}{\partial n}\Bigr|_{\substack{n_{gs}}(\mathbf{r})}\delta n(\mathbf{r},\omega)\label{eq:1}
\end{equation}

In Eq.~(\ref{eq:1}) $\delta n(\mathbf{r},\omega)$ is the induced
density and the adiabatic local density
approximation~\cite{GROSS1990255} (ALDA) to the exchange-correlation
kernel has been introduced. The induced density is obtained from the
Kohn--Sham (non-interacting) dielectric susceptibility,
$\chi(\mathbf{r},\mathbf{r}^{\prime},\omega)$ and the SCF potential
as:

\begin{equation}
\delta n(\mathbf{r},\omega)=\int\chi(\mathbf{r},\mathbf{r}^{\prime},\omega)v^{\textit{SCF}}(\mathbf{r}^{\prime},\omega)d\mathbf{r}^{\prime}
\label{eq:2}
\end{equation}
Substitution of Eq.~(\ref{eq:2}) in Eq.~(\ref{eq:1}) gives the equation:

\begin{equation}
v^{\textit{SCF}}(\mathbf{r},\omega)=v^{\textit{ext}}(\mathbf{r},\omega)+\int\int K(\mathbf{r},\mathbf{r}^{\prime})\chi(\mathbf{r}^{\prime},\mathbf{r}^{\prime\prime}\omega)v^{\textit{SCF}}(\mathbf{r}^{\prime\prime}\omega)d\mathbf{r}^{\prime}d\mathbf{r}^{\prime\prime}
\label{eq:3}
\end{equation}

where in Eq.~(\ref{eq:3}) $K(\mathbf{r},\mathbf{r}^{\prime})$ is the Coulomb and the ALDA exchange-correlation kernel:
\begin{equation}
K(\mathbf{r},\mathbf{r}^{\prime})=\frac{1}{|\mathbf{r}-\mathbf{r}^{\prime}|}+\delta(\mathbf{r}-\mathbf{r}^{\prime})\frac{\partial
  v_{xc}}{\partial n}\Bigr|_{\substack{n_{gs}}(\mathbf{r})}.
\label{eq:4}
\end{equation}
The integral equation, Eq.~(\ref{eq:3}), is solved with respect to
$v^{\rm SCF}(\mathbf r,\omega)$:
representing the equation with the same B-spline LCAO basis set leads to the solution of a linear system of algebraic equations. An efficient way to implement the construction of the 
 matrix in the present B-spline basis set has been described previously~\cite{jcp_1.1937367}. Once $v^{\textit{SCF}}$ is obtained, the total cross section is calculated via transition moments using $v^{\textit{SCF}}$ instead of the dipole operator.

The TDDFT calculations have been carried out by including the set of
dipole-allowed excitation channels arising from all orbitals with
$|\varepsilon_i^{KS}|<$7.0~a.u. while deeper orbitals have been kept
frozen. Convergence of the computed total cross sections and asymmetry
parameters for the excitation energies considered in the present work
has been tested in preliminary calculations.




As a further convergence test directly relevant to the phase-sensitive
observable, we repeated the CH$_3$I calculation near 105.9~eV with the
single-center cutoff increased from $l_{\max}=12$ to 16 and, subsequently,
with the radial box enlarged from $R_{\max}=25$ to 35~a.u. Increasing
$l_{\max}$ changed the corrected polarization-axis EWS delay from
30.63 to 30.83~as, while the combined $l_{\max}=16$,
$R_{\max}=35$ calculation gave approximately 30.8~as. The corresponding
cross section changed by less than 0.2\%, while the change in $\beta$
remained below 0.01. In the radial-box test the independently determined
LB94 matching shift changed from 5.67281225 to 5.19307038~bohr; using the
appropriate shift for each box, as prescribed in Sec.~II\,D, removes an
otherwise artificial $\sim6$~as phase contribution. These tests confirm
that the reported CH$_3$I EWS delay is well converged with respect to both
the angular cutoff and radial-box size.

\subsection{Molecular EWS delay}

Molecular photoionization time delays have been addressed within a
number of theoretical approaches
\cite{Chacon2014,Hockett2016,Baykusheva2017,Serov2017,Borras2023,SunerRubio2024,GonzalezCollado2024,Rescigno2024,Pranjal2025}.
In the present work, for a fixed molecular orientation and
photoelectron emission direction, let the one-photon transition
amplitude be
\begin{equation}
 A(E,\hat{\bm k})=|A(E,\hat{\bm k})|
 \exp[i\phi(E,\hat{\bm k})].
\end{equation}
The corresponding EWS delay may be expressed
\cite{Pazourek2015,Kheifets2013} in a numerically stable bilinear
form
\begin{equation}
 \tau(E,\hat{\bm k})=
 \frac{\operatorname{Im}\left[
 A^*(E,\hat{\bm k})\,\partial_E A(E,\hat{\bm k})
 \right]}
 {|A(E,\hat{\bm k})|^2}.
 \label{eq:bilinear}
\end{equation}
Away from zeros of the amplitude this is identical to
$\partial_E\arg A$.  Equation~(\ref{eq:bilinear}) is preferable in a
multichannel calculation because it avoids explicit phase unwrapping and
makes the weighting by the observable intensity automatic.

For a randomly oriented target ionized by linearly polarized light, the
standard Racah reduction leaves only the scalar and quadrupole tensor
components.  We therefore write
\begin{align}
 S(E,\theta) &= \frac{1}{4\pi}
 \left[S_0(E)+S_2(E)P_2(\cos\theta)\right],\\
 N(E,\theta) &= \frac{1}{4\pi}
 \left[N_0(E)+N_2(E)P_2(\cos\theta)\right],
\end{align}
where
\begin{equation}
 S=|A|^2,\qquad
 N=\operatorname{Im}\left[A^*\partial_E A\right].
\end{equation}
The angular-momentum content of these coefficients is most transparent in
the corresponding one-center reduction.  Absorbing the scattering phase
and reduced radial matrix element into the complex dipole $D_\ell(E)$, with
$\ell=\ell_i\pm1$ for electric-dipole ionization and
$\hat{\ell}\equiv\sqrt{2\ell+1}$, the isotropic ($K=0$) coefficients are
\begin{align}
 S_0(E)
 &= \frac{1}{3}
 \sum_{\ell=\ell_i\pm1}|D_\ell(E)|^2,
 \label{eq:S0}\\
 N_0(E)
 &= \frac{1}{3}
 \sum_{\ell=\ell_i\pm1}
 \operatorname{Im}\!\left[
 D_\ell^*(E)\,\partial_E D_\ell(E)
 \right].
 \label{eq:N0}
\end{align}
The quadrupole ($K=2$) coefficients contain the interference between the
allowed continuum partial waves,
\begin{align}
 S_2(E)
 &= \sqrt{\frac{10}{3}}
 \sum_{\ell,\ell'=\ell_i\pm1}
 \hat{\ell}\,\hat{\ell}'\,
 (-1)^{\ell_i+\ell+\ell'}
 \nonumber\\
 &\quad\times
 \begin{pmatrix}
  \ell & \ell' & 2\\
  0 & 0 & 0
 \end{pmatrix}
 \left\{
 \begin{matrix}
  2 & 1 & 1\\
  \ell_i & \ell & \ell'
 \end{matrix}
 \right\}
 D_\ell(E)D_{\ell'}^*(E),
 \label{eq:S2}\\[1ex]
 N_2(E)
 &= \sqrt{\frac{10}{3}}
 \sum_{\ell,\ell'=\ell_i\pm1}
 \hat{\ell}\,\hat{\ell}'\,
 (-1)^{\ell_i+\ell+\ell'}
 \nonumber\\
 &\quad\times
 \begin{pmatrix}
  \ell & \ell' & 2\\
  0 & 0 & 0
 \end{pmatrix}
 \left\{
 \begin{matrix}
  2 & 1 & 1\\
  \ell_i & \ell & \ell'
 \end{matrix}
 \right\}
 \nonumber\\
 &\quad\times
 \operatorname{Im}\!\left[
 D_\ell^*(E)\,\partial_E D_{\ell'}(E)
 \right].
 \label{eq:N2}
\end{align}
Thus the delay numerator has exactly the same Racah coupling structure as
the photoelectron angular distribution: one dipole factor is replaced by
its energy derivative.  In the same normalization $S_0$ is proportional to
the total cross section and $S_2/S_0=\beta$.  The full multicenter molecular
expressions retain the additional molecular-frame and channel indices, but
the rank-0/rank-2 contraction structure is unchanged.

The observable delay is therefore
\begin{equation}
 \tau(E,\theta)=
 \frac{N_0(E)+N_2(E)P_2(\cos\theta)}
 {S_0(E)+S_2(E)P_2(\cos\theta)}.
 \label{eq:tensor_delay}
\end{equation}
To connect this notation explicitly with the tensor contractions used in
our numerical implementation, we introduce a common normalization factor
$C(E)$, which cancels from the final delay, and define
\begin{align}
 S_0(E) &= C(E)a_0(E),
 &S_2(E) &= C(E)\sqrt{30}\,a_2(E),
 \nonumber\\
 N_0(E) &= C(E)\operatorname{Im}b_0(E),
 &N_2(E) &= C(E)\sqrt{30}\,\operatorname{Im}b_2(E).
 \label{eq:ab_mapping}
\end{align}
Here $a_K$ denotes the ordinary dipole--dipole tensor contraction and
$b_K$ the corresponding energy-derivative contraction.  Consequently,
\begin{equation}
 \beta(E)=\frac{S_2(E)}{S_0(E)}
 =\sqrt{30}\,\frac{a_2(E)}{a_0(E)}.
 \label{eq:beta_ab}
\end{equation}
The factor $\sqrt{30}$ is therefore not an additional dynamical factor; it
is the conventional normalization of the rank-2 Cherepkov tensor
contraction generated by the $K=2$ angular-momentum coupling in
Eqs.~(\ref{eq:S2}) and (\ref{eq:N2}).  Along the polarization axis,
$P_2(1)=1$, so that $S_0+S_2\propto\sigma(E)[1+\beta(E)]$ and the
implementation evaluates
\begin{equation}
 \tau_{\parallel}(E)=
 \frac{\operatorname{Im}[b_0(E)+\sqrt{30}\,b_2(E)]}
 {a_0(E)[1+\beta(E)]}.
 \label{eq:implementation}
\end{equation}
This construction is the molecular counterpart of the angular-momentum
formalism used previously for atomic time delays \cite{Kheifets2020}.

The energy derivative of each complex dipole was obtained by five-point
finite differences on a photon-energy grid with a spacing of $0.1$~a.u.
As consistency checks, an $s$-shell
calculation, for which only the $p$ wave contributes, recovers $\beta=2$ and
an angle-independent delay $\tau=\partial_E\arg D_1$.  More generally, the
implementation gives angle independence whenever a single continuum partial
wave is retained and agrees with the direct phase-derivative definition in
the corresponding atomic tests.

\subsection{TDDFT photoionization amplitudes}

The molecular calculations employ the B-spline multicenter TDDFT
framework developed by \citet{Toffoli2002} and \citet{Fukuzawa2019}.
The Kohn--Sham ground state and the linear response to the dipole
field are described with the van Leeuwen--Baerends (LB94)
exchange-correlation potential \cite{PhysRevA.49.2421}, whose
correct Coulombic asymptotic behavior is essential for continuum
photoionization.  The TDDFT response produces a screened dipole field
and complex continuum dipoles that include interchannel coupling at
the linear-response level.

Calculations were performed for CH$_3$I, CH$_3$CH$_2$I,
CH$_3$CHICH$_3$, and two conformers of CH$_3$CH$_2$CH$_2$I,
including the lowest-energy and a slightly higher eclipsed geometry.
The present theoretical comparison is restricted to these molecules,
for which complete TDDFT photoionization amplitudes are available;
the two iodobutane isomers are included only in the experimental
overview
of \Fref{fig:experiment}.

The iodine $4d$ manifold in CH$_3$I provides a useful illustration.
In the underlying molecular-orbital representation the five
$4d$-derived components are $10a_1$ and the two doubly degenerate $6e$
and $7e$ levels, with orbital energies clustered near $-57$ eV. The
present molecular TDDFT treatment does not explicitly resolve the
iodine $4d_{5/2}$--$4d_{3/2}$ spin--orbit splitting. The small
splitting of the molecular components compared with the width of the
giant resonance already suggests that the XUV step should retain a
strongly atomic character.

The final complex dipoles are $S$-matrix normalized.  In the continuum
machinery a real standing-wave $K$ matrix is transformed to outgoing-wave
normalization according to
\begin{equation}
 {\bm D}^{(+)}=(\bm I-i\bm K)^{-1}{\bm D}.
 \label{eq:Ktransform}
\end{equation}
For an eigenchannel, $K_\alpha=\tan\delta_\alpha$ gives
$(1-iK_\alpha)^{-1}=\cos\delta_\alpha\exp(i\delta_\alpha)$, explicitly
placing the physical continuum scattering phase into the photoionization
amplitude.

\subsection{Asymptotic LB94 phase convention}
\label{sec:slb}

The time-delay calculation exposed a feature of the continuum matching that
is invisible to the cross section and $\beta$.  The LB94 potential is
matched to a Coulomb tail using a shifted radial coordinate,
\begin{equation}
 \rho=k(r+S_{\rm LB94}),
 \label{eq:shiftedrho}
\end{equation}
rather than simply $\rho=kr$.  The shift is fixed by continuity of the
asymptotic potential at the fitting radius,
\begin{equation}
 S_{\rm LB94}=-\frac{1}{V_{\rm LB94}(R_{\rm fit})}-R_{\rm fit}.
 \label{eq:slbdef}
\end{equation}
For CH$_3$I, with $R_{\max}=25$ a.u., we obtain $S_{\rm LB94}=5.67281225$ bohr.

At the simplest asymptotic level,
\begin{equation}
 \exp(ikr)\longrightarrow
 \exp[ik(r+S_{\rm LB94})]
 =\exp(ik S_{\rm LB94})\exp(ikr),
\end{equation}
so that the shifted reference functions carry an additional smooth phase
\begin{equation}
 \phi_s(E)\simeq k(E)S_{\rm LB94},
 \qquad
 \frac{d\phi_s}{dE}\simeq \frac{S_{\rm LB94}}{k(E)}.
 \label{eq:phaseshift}
\end{equation}
For the relevant photoelectron energies this derivative is tens of
attoseconds and therefore cannot be ignored in a phase-sensitive
observable.

The key diagnostic is that a common factor of unit modulus,
\begin{equation}
 \widetilde D_i(E)=D_i(E)\exp[i k(E)S_{\rm LB94}],
 \label{eq:rephase}
\end{equation}
leaves all products $D_iD_j^*$ unchanged.  Consequently the cross
section and $\beta$ parameter are strictly invariant, whereas the
Wigner delay is shifted by the derivative of the common phase.
Applying Eq.~(\ref{eq:rephase}) to the already computed $S$-matrix
dipoles removes the large smooth background that otherwise persists at
high energy.

\begin{figure}[t]
\centering
\includegraphics[width=\columnwidth]{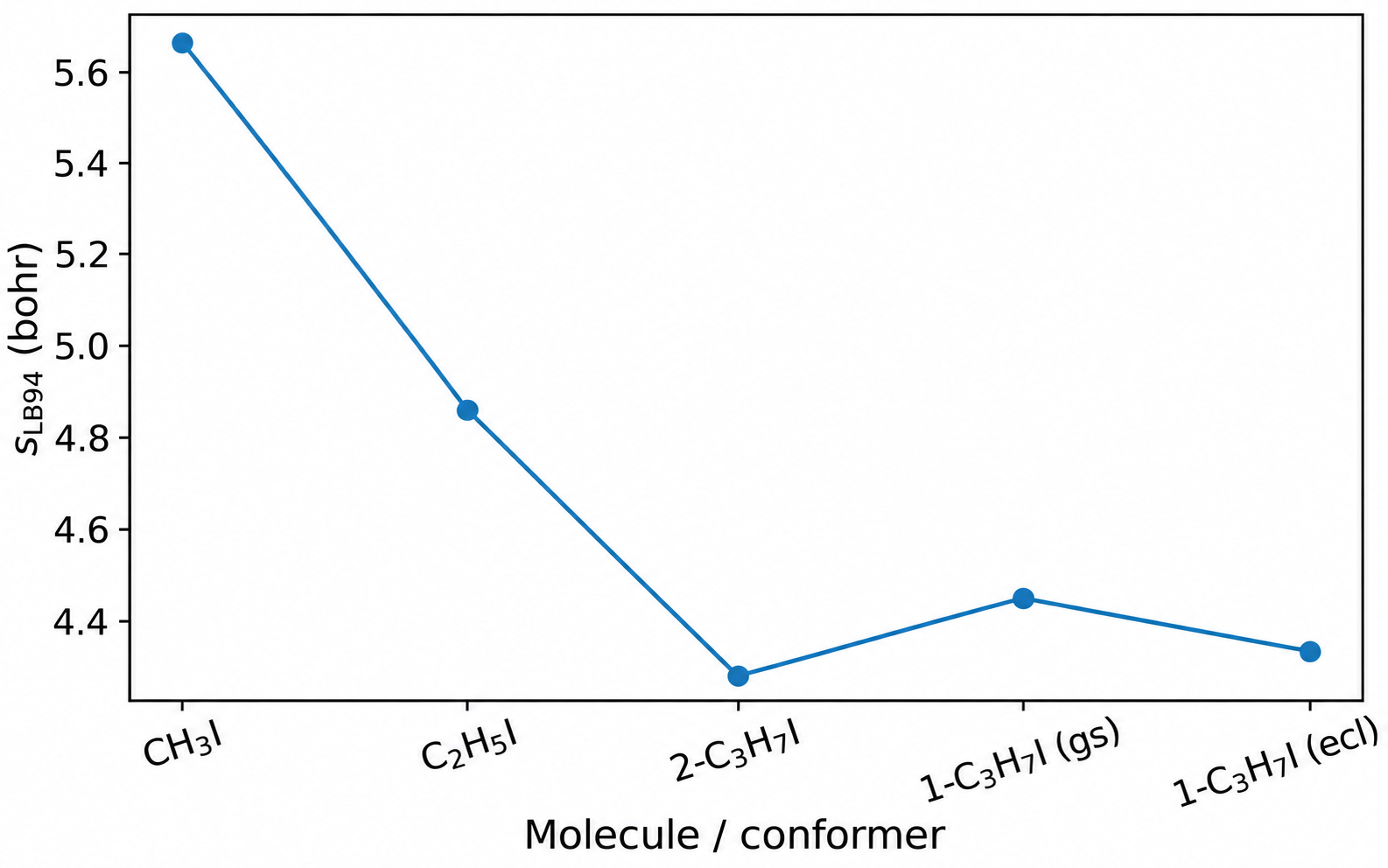}
\caption{
LB94 radial matching shift $S_{\rm LB94}$ for the calculated
iodoalkanes. These shifts define the molecule-dependent asymptotic
phase
correction of Eq.~(23) and are fixed independently by the continuum
matching.
}
\label{fig:slbshifts}
\end{figure}

The values of $S_{\rm LB94}$ for the molecular sequence are shown in
\Fref{fig:slbshifts}.  They quantify the radial displacement entering
the asymptotic matching and therefore set the scale of the phase
correction in Eq.~(\ref{eq:rephase}).  Because these shifts are fixed
independently by the asymptotic matching, the associated phase
correction is not an adjustable molecular contribution.  Applying the
correction removes the convention-dependent background but does not
generate the experimental molecular ordering.

\begin{table}[t]
\caption{LB94 asymptotic radial shifts used in the molecular continuum
matching with $R_{\max}=25$ a.u.}
\label{tab:slb}
\begin{ruledtabular}
\begin{tabular}{lc}
System & $S_{\rm LB94}$ (bohr)\\
\hline
CH$_3$I & 5.67281225\\
CH$_3$CH$_2$I & 4.85177588\\
CH$_3$CHICH$_3$ & 4.28604993\\
CH$_3$CH$_2$CH$_2$I, ground conformer & 4.44010296\\
CH$_3$CH$_2$CH$_2$I, eclipsed conformer & 4.33444225\\
\end{tabular}
\end{ruledtabular}
\end{table}

\begin{figure}[t]
\centering
\includegraphics[width=\columnwidth]{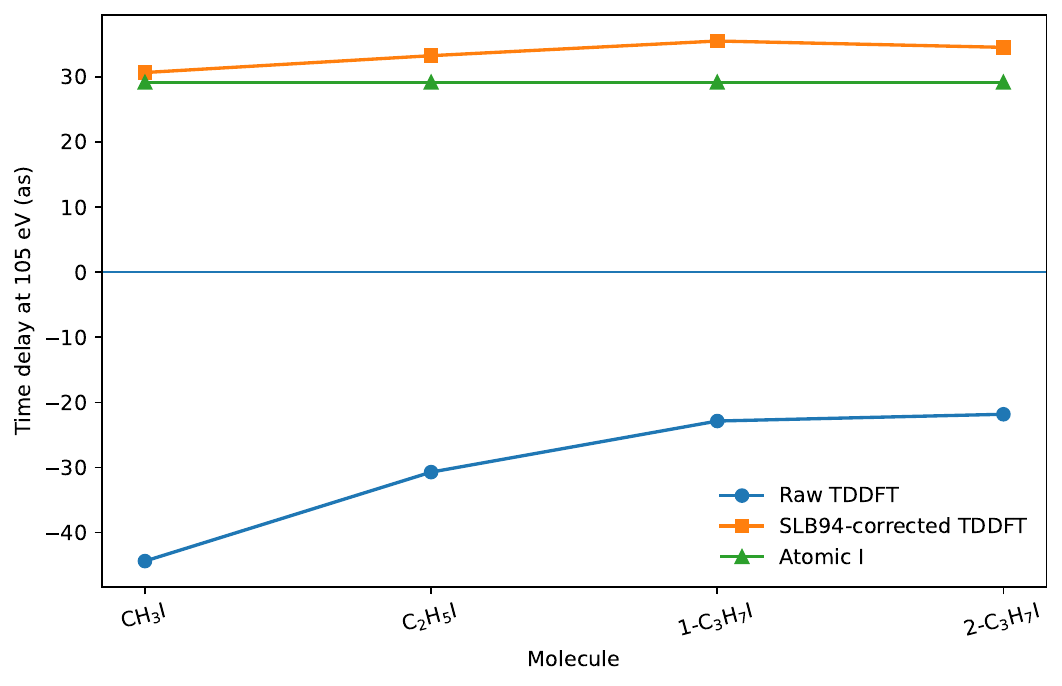}
\caption{Effect of the LB94 asymptotic phase correction at 105~eV.  The raw
TDDFT delays contain a large convention-dependent background.  Rephasing
the $S$-matrix-normalized dipoles by the LB94 matching shift introduced in
\Fref{fig:slbshifts} restores the atomic-I scale.  The cross section and
$\beta$ are unchanged by this unit-modulus transformation.}
\label{fig:slbcorrection}
\end{figure}

The numerical effect of applying this correction is demonstrated in
\Fref{fig:slbcorrection} at 105~eV for the four molecules used in the
direct comparison with experiment.  Before rephasing, the molecular
delays contain a large negative, convention-dependent offset.  The
LB94 phase correction, determined independently by the asymptotic
matching, moves all four values into the 30--36 as range, close to the
atomic-I scale, while leaving the cross section and $\beta$ unchanged.
This provides a direct numerical diagnostic that the smooth background
originates from the asymptotic continuum phase convention rather than
from a failure of the LB94 potential itself.

\section{Atomic iodine benchmarks}
\label{sec:atomic}

Before interpreting the molecular calculations it is useful to
establish the natural scale and robustness of the iodine $4d$ EWS
delay.  \citet{PiLandsman2018} calculated neutral atomic iodine using
time-dependent local-density approximation (TDLDA).  Across the broad
$4d\rightarrow\epsilon f$ giant resonance, approximately 60--160~eV in
photon energy, the dominant continuum phase changes smoothly by about
$\pi$.  A simple estimate from this phase excursion gives a delay of
order 20 as, and the full calculation gives about 30 as near 100~eV.
A negative delay structure appears at the higher-energy Cooper
minimum.

The iodide anion provides a complementary reference because
photodetachment leaves neutral iodine and therefore removes the
asymptotic Coulomb phase.  \citet{Grafstrom2023} performed a
33-channel RRPA calculation for I$^-$; \citet{Banerjee2021} calculated
angle-resolved I$^-$ delays including orbital relaxation.  The
threshold region is qualitatively different from neutral-I
photoionization, as expected, but once the photoelectron is well
inside the giant-resonance region the calculated $4d$ delays become
remarkably similar.

\begin{figure}[t]
\centering
\includegraphics[width=\columnwidth]{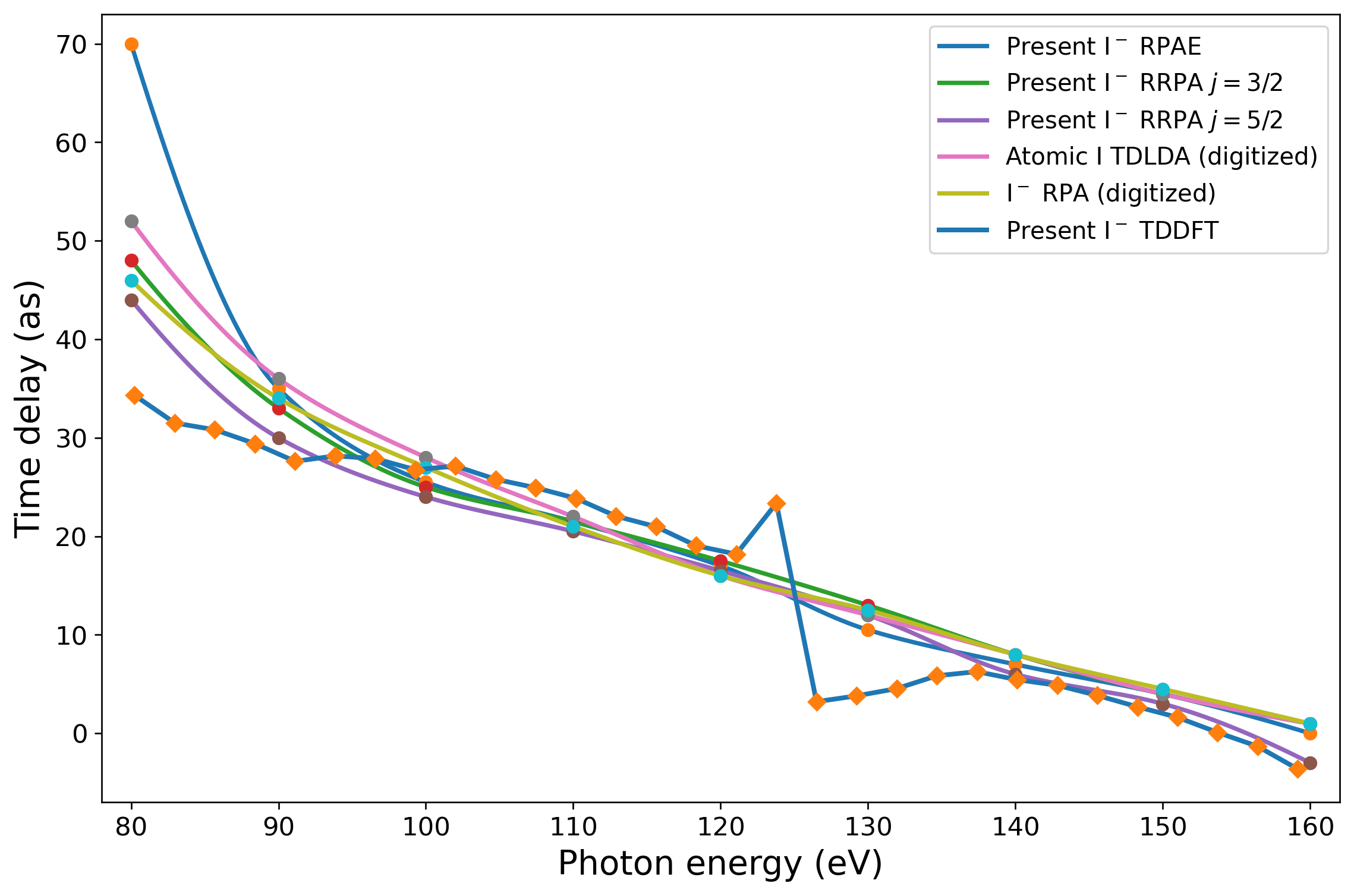}
\caption{ Comparison of atomic iodine $4d$ delays in the
  giant-resonance region. Present I$^-$ results obtained within the
  non-relativistic RPAE \cite{Kheifets2013} and relativistic RRPA
  \cite{PhysRevA.96.053407} models, together with the present
  molecular-symmetry TDDFT calculation for I$^-$, are overplotted with
  values digitized from the neutral-I TDLDA calculation of
  \citet{PiLandsman2018} and the I$^-$ RRPA calculation of
  \citet{Grafstrom2023} .}
\label{fig:atomic}
\vs{-0.5cm}
\end{figure}

\Fref{fig:atomic} makes this comparison explicit.  The literature
curves have been digitized from the source plots and are included
to emphasize the scale and common trend.  The present
molecular-symmetry TDDFT calculation for I$^-$ provides an independent
benchmark and, after the LB94 asymptotic phase correction, closely
follows the atomic calculations through the central part of the giant
resonance.  Away from threshold, neutral I, I$^-$, RPAE, relativistic
RRPA, and TDDFT all cluster together.  Around 90--120~eV their spread
is only of order several attoseconds to, at most, about ten
attoseconds.

This clustering is not a trivial consequence of identical asymptotic
potentials: neutral I and I$^-$ have different long-range behavior close to
threshold.  Rather, it demonstrates that the broad giant resonance is
governed predominantly by a robust, local iodine-centered continuum
response.  This observation will become central when the molecular sequence
is considered.

\section{Iodoalkane results}
\label{sec:results}

\subsection{Cross section and angular distribution}

The first test of the molecular TDDFT description is provided by
phase-insensitive observables.  Across the calculated sequence the I $4d$
cross sections retain the broad giant-resonance profile and the averaged
anisotropy parameters are very similar.  The weak molecular dependence of
$\beta$ was already apparent in the calculations: the ionized
level remains essentially atomic, and changing the alkyl group has only a
modest influence on the dominant $4d\rightarrow\epsilon f$ continuum.

A complementary check is provided by the synchrotron
photoionization measurements of \citet{Forster2026} for
CH$_3$I, C$_2$H$_5$I, 1-C$_3$H$_7$I, and 2-C$_3$H$_7$I.
The experimental cross sections were extracted from the
relative photoelectron yields measured at the magic angle
and normalized to the CH$_3$I absorption data of
\citet{OSullivan1982}. They are therefore used primarily
to compare the spectral profiles rather than the absolute
normalization. For the theory, we use the molecule-specific
TDDFT cross sections extracted from the same calculations
as the time-delay analysis. A single common multiplicative
normalization, fixed by CH$_3$I, is applied to all TDDFT
curves so that their relative molecular variation is preserved.

\begin{figure}[t]
\centering
\includegraphics[width=0.5\textwidth]{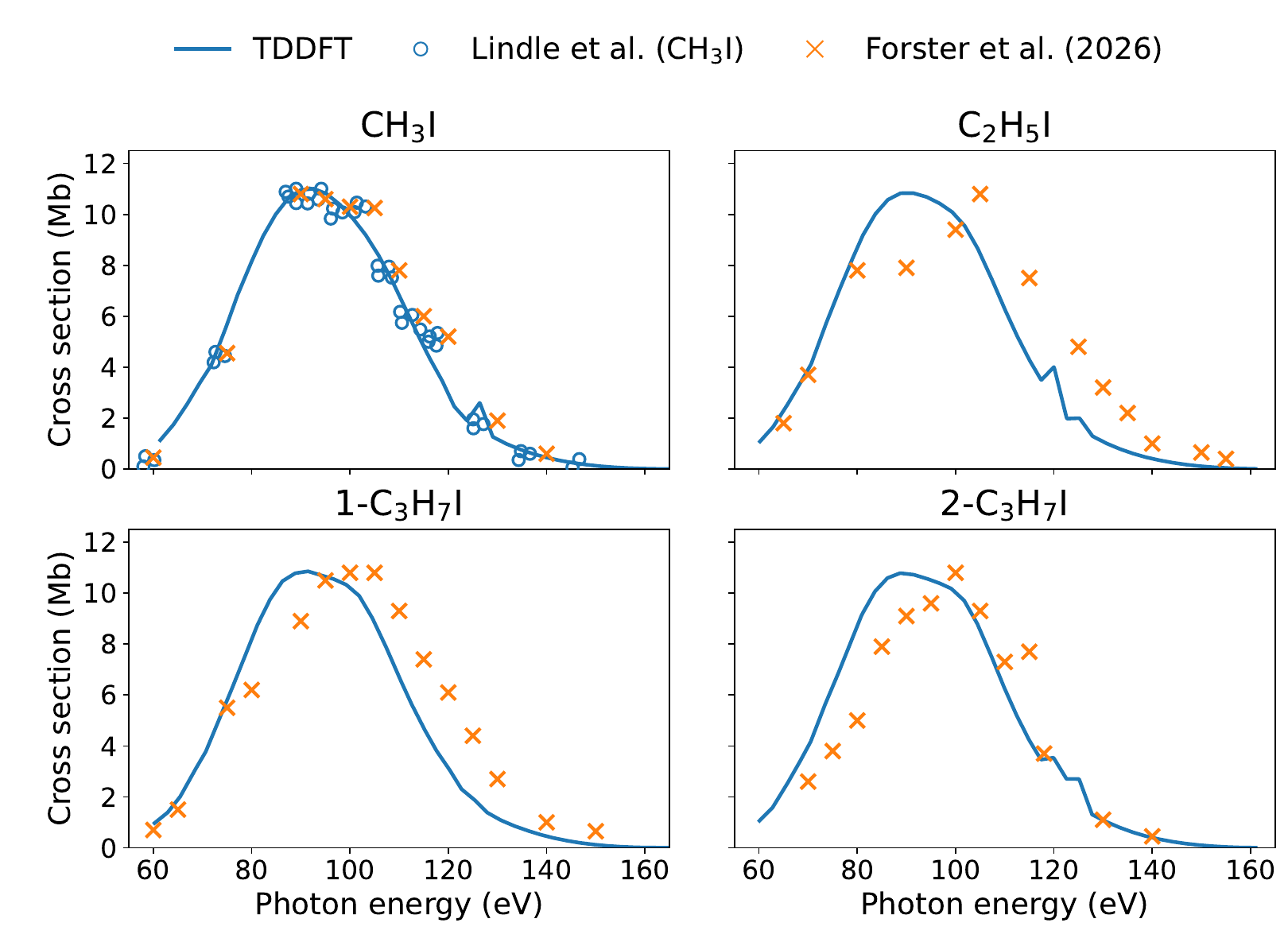}
\caption{Comparison of the I $4d$ giant-resonance cross sections for
  CH$_3$I, C$_2$H$_5$I, 1-C$_3$H$_7$I, and 2-C$_3$H$_7$I. Solid
  curves show the present molecule-specific TDDFT results, and crosses
  show the synchrotron measurements of \citet{Forster2026}.
  In the CH$_3$I panel, open circles show the independent experimental
  cross section of \citet{Lindle1984}.
  The synchrotron data of \citet{Forster2026} were normalized to the
  CH$_3$I absorption measurements of \citet{OSullivan1982}.
  A single common normalization factor, fixed by CH$_3$I, is applied
  to all TDDFT curves, preserving their relative molecular dependence.
  The comparison thus primarily tests the giant-resonance profiles
  rather than the absolute oscillator strengths.}\label{fig:crosssections}
\end{figure}

\Fref{fig:crosssections} shows that the broad I $4d$ giant-resonance
envelope
is largely preserved across the molecular sequence. In CH$_3$I the two
experimental data sets follow the same overall resonance profile, and
the TDDFT curve reproduces the broad rise, maximum, and decay of the
cross section. For the longer iodoalkanes, the synchrotron
measurements
of \citet{Forster2026} show some redistribution on the rising and
falling sides of the resonance, but no qualitatively different
profile.
In particular, the two propyl isomers remain rather similar through
the main 90--110~eV resonance region despite the much larger isomer
dependence reported for the streaking delay. Given the experimental
normalization, this comparison should not be interpreted as a precise
test of absolute oscillator strength. It does, however, reinforce the
conclusion that the gross one-photon I $4d$ response is only weakly
modified by the molecular environment.

\begin{figure}[t]
\centering
\includegraphics[width=1.04\columnwidth]{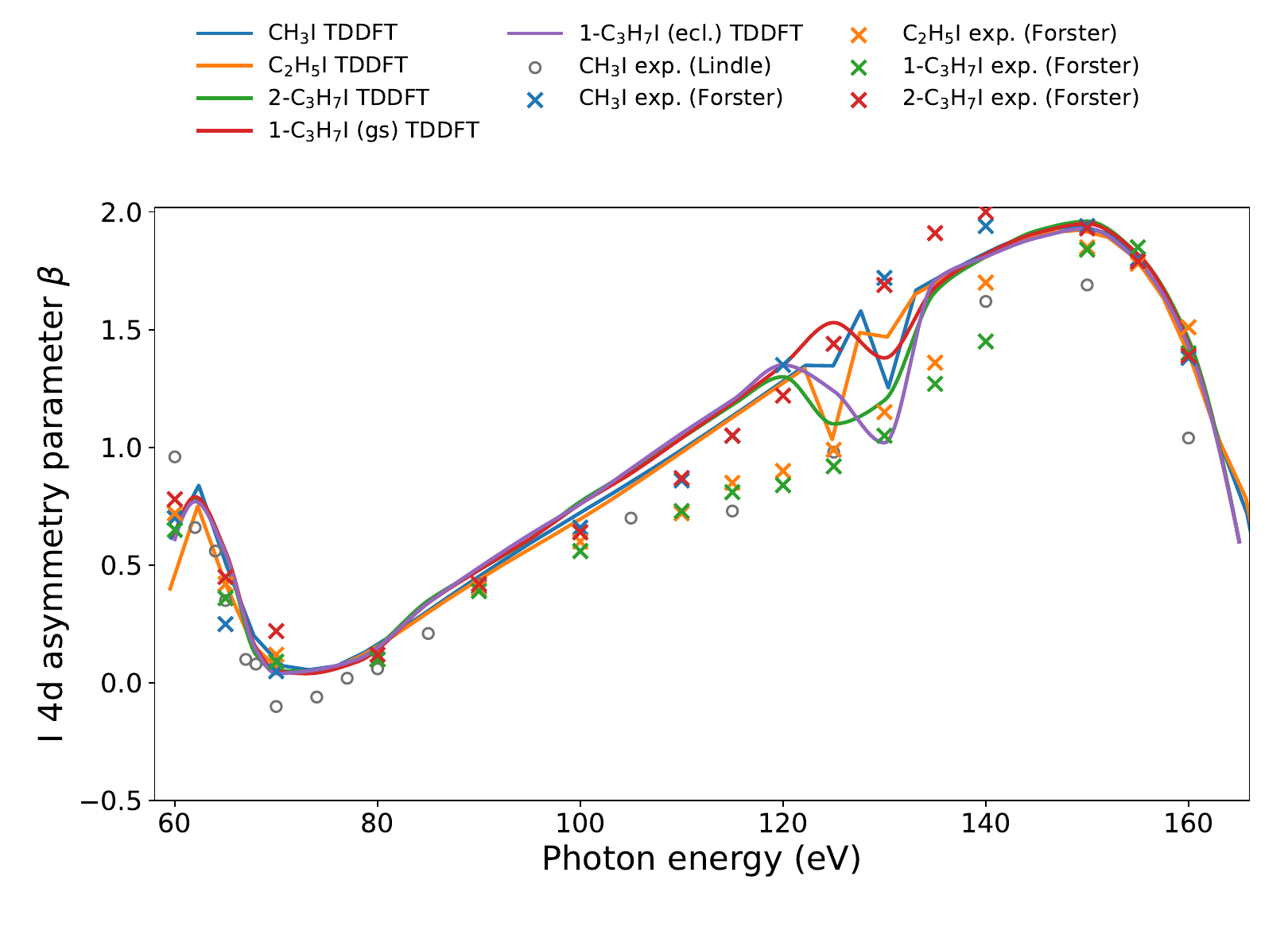}
\caption{I $4d$ photoelectron asymmetry parameter $\beta$ for CH$_3$I,
C$_2$H$_5$I, 2-C$_3$H$_7$I, and the ground-state (gs) and eclipsed
(ecl)
geometries of 1-C$_3$H$_7$I.
Solid curves show the present TDDFT results. Crosses show the
synchrotron measurements of \citet{Forster2026}, while open circles
represent the independent CH$_3$I measurements of \citet{Lindle1984}.
The experimental points were digitized from the respective
source figures.}
\label{fig:beta}
\end{figure}

The calculated anisotropy parameters are compared with the
experimental measurements of \citet{Forster2026} in
\Fref{fig:beta}. All five TDDFT molecular/conformer curves display
essentially the same global energy dependence: a broad minimum
around 70--75~eV, followed by a steady rise toward $\beta\simeq 2$
near 145--150~eV and a rapid decrease at still higher photon
energy. The largest separation among the calculated molecular
curves is confined to the 120--135~eV region, where relatively
narrow structure is superimposed on the common trend. The
experimental measurements follow the same overall evolution but
show a more pronounced molecular dependence between 110 and
140~eV, particularly for C$_2$H$_5$I and 1-C$_3$H$_7$I.

The overall consistency in \Fref{fig:beta} supports the one-photon
molecular description.  In particular, the angular distribution does
not show a systematic molecular variation comparable in scale or
ordering to the pronounced differences observed in the streaking
delays.  Thus the delay is not being extracted from a continuum whose
amplitude or angular distribution changes dramatically from one
molecule to another.  The field-free molecular environment certainly
causes scattering and partial-wave mixing, but it does not
substantially reconstruct the giant resonance.  This contrasts with
HF+LDA scattering calculations discussed in Ref.~\cite{Schroder2025},
where the giant resonance itself is not rendered adequately.

\subsection{Corrected molecular EWS delays}

\begin{figure}[t]
\centering
\includegraphics[width=\columnwidth]{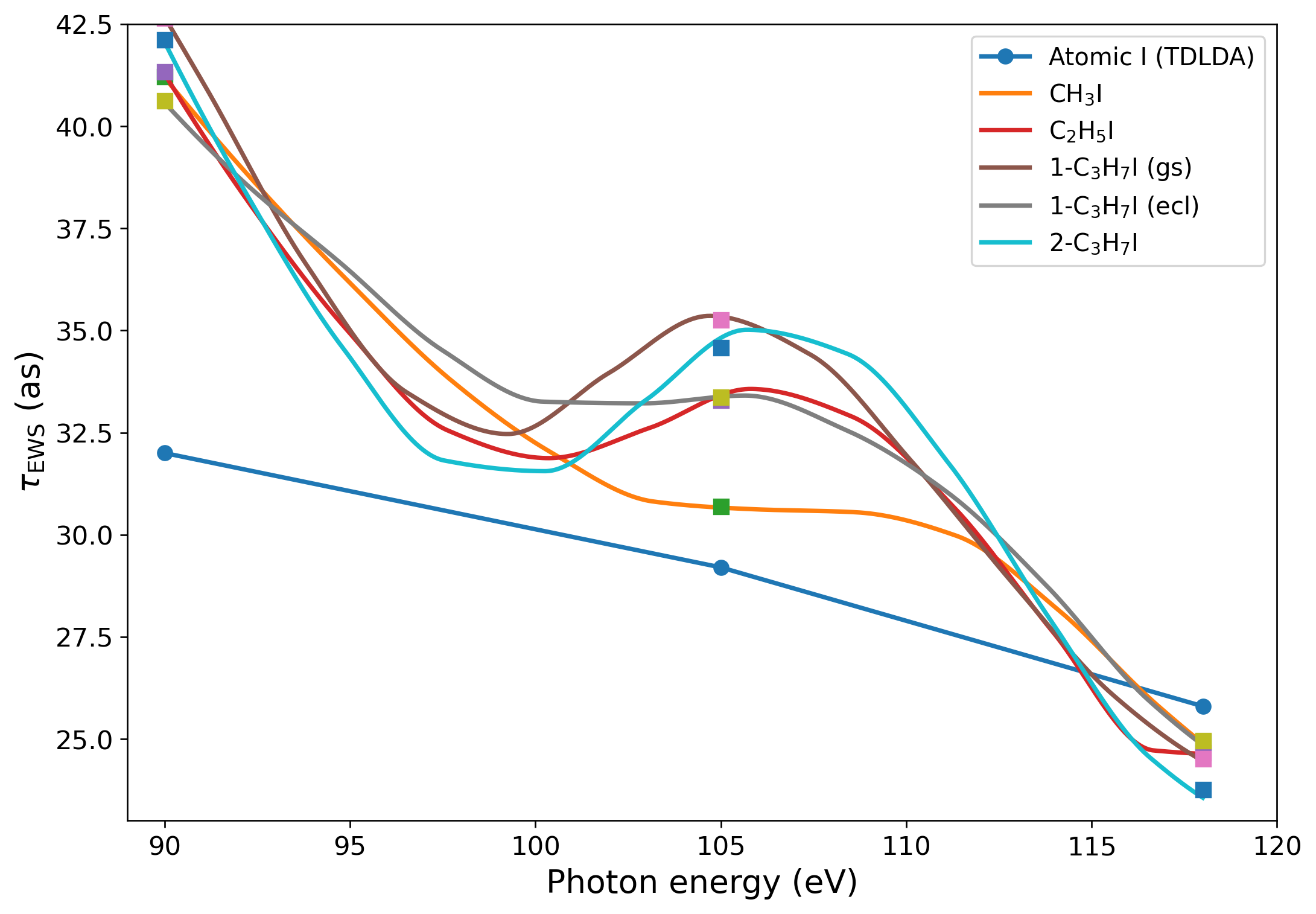}
\caption{$S_{\rm LB94}$-corrected field-free TDDFT EWS delays for
CH$_3$I, C$_2$H$_5$I, the ground-state and eclipsed conformers of
1-iodopropane, and 2-iodopropane compared with the atomic-I TDLDA
reference. }
\label{fig:bundle}
\end{figure}

\Fref{fig:bundle} shows the corrected TDDFT delays along the polarization axis
continuously over the 90--118~eV experimental energy range. The five
molecular/conformer curves---CH$_3$I, C$_2$H$_5$I, 2-iodopropane,
and the ground-state and eclipsed conformers of 1-iodopropane---form
a narrow bundle around the atomic-I TDLDA reference. The symbols mark
the values at the three experimental photon energies. The continuous
curves show that this close clustering is not confined to the three
sampled energies but persists throughout the experimental range.

At the three experimental photon energies, all five
molecular/conformer curves remain close to the atomic-I scale. For the
four molecules for which direct experimental comparisons are
available, the numerical values are collected in
Table~\ref{tab:comparison}. At 118~eV all four calculated molecular
delays are essentially identical to the atomic value within about
2~as. At 105~eV the molecular shifts are between about 1.5 and
6.3~as. Even at 90~eV, where the largest field-free molecular
correction occurs, all four results remain within about 10~as of
atomic iodine.

\begin{table*}[t]
\caption{Atomic-I reference, corrected TDDFT delays for
  emission along the polarization axis, and experimental
  streaking-derived delays, in attoseconds.  The experimental values
  are from Ref.~\cite{Schroder2025}.
}
\label{tab:comparison}
\begin{ruledtabular}
\begin{tabular}{lcccccc}
 & \multicolumn{2}{c}{90~eV} & \multicolumn{2}{c}{105~eV} &
 \multicolumn{2}{c}{118~eV}\\
System & TDDFT & Experiment & TDDFT & Experiment & TDDFT & Experiment\\
\hline
Atomic I & 32.0 & -- & 29.2 & -- & 25.8 & --\\
CH$_3$I & 41.2 & $76.9^{+5.3}_{-11.6}$ & 30.7 & $40.9\pm4.1$ &
24.9 & $40.1\pm4.0$\\
C$_2$H$_5$I & 41.3 & $62.8^{+6.2}_{-12.5}$ & 33.3 & $39.6\pm7.1$ &
24.7 & $25.1\pm3.3$\\
1-C$_3$H$_7$I & 38.8 & $36.6^{+4.1}_{-10.5}$ & 35.5 & $23.1\pm7.2$ &
25.6 & $10.3\pm5.6$\\
2-C$_3$H$_7$I & 42.1 & $78.5^{+6.7}_{-13.0}$ & 34.6 & $63.9\pm7.1$ &
23.8 & $20.0\pm5.7$\\
\end{tabular}
\end{ruledtabular}
\end{table*}

\begin{figure}[t]
\centering
\includegraphics[width=\columnwidth]{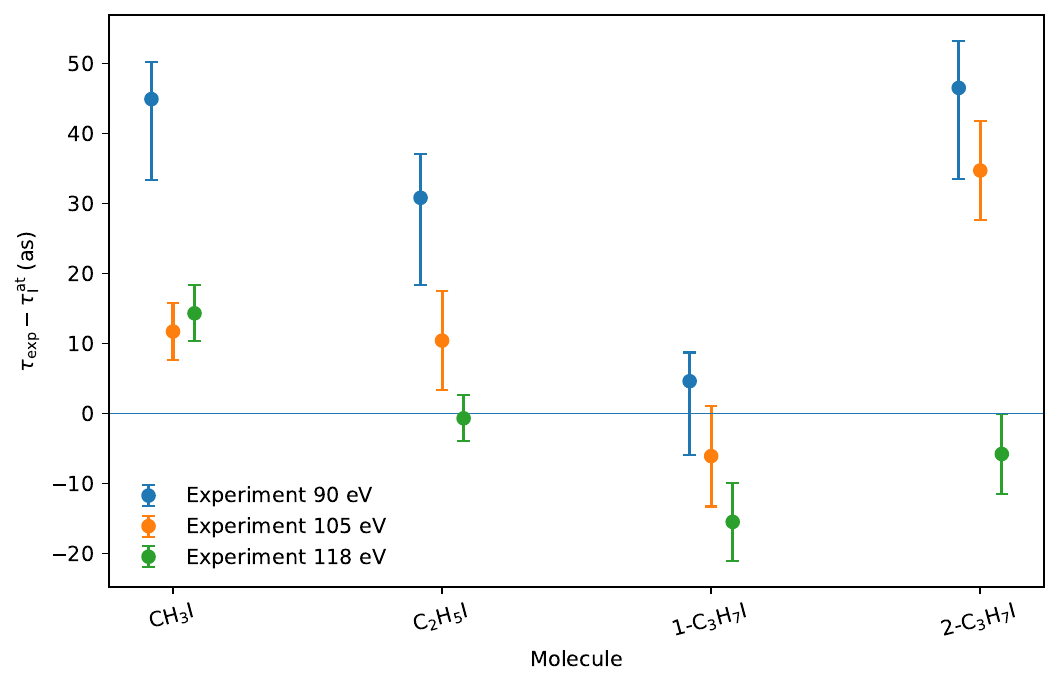}
\caption{Experimental molecular departure from the atomic-I TDLDA delay for
the four molecules for which direct TDDFT comparisons are presently
available.  The deviations are large, nonmonotonic, and strongly
isomer-dependent.}
\label{fig:expminus}
\end{figure}

\begin{figure}[t]
\centering
\includegraphics[width=\columnwidth]{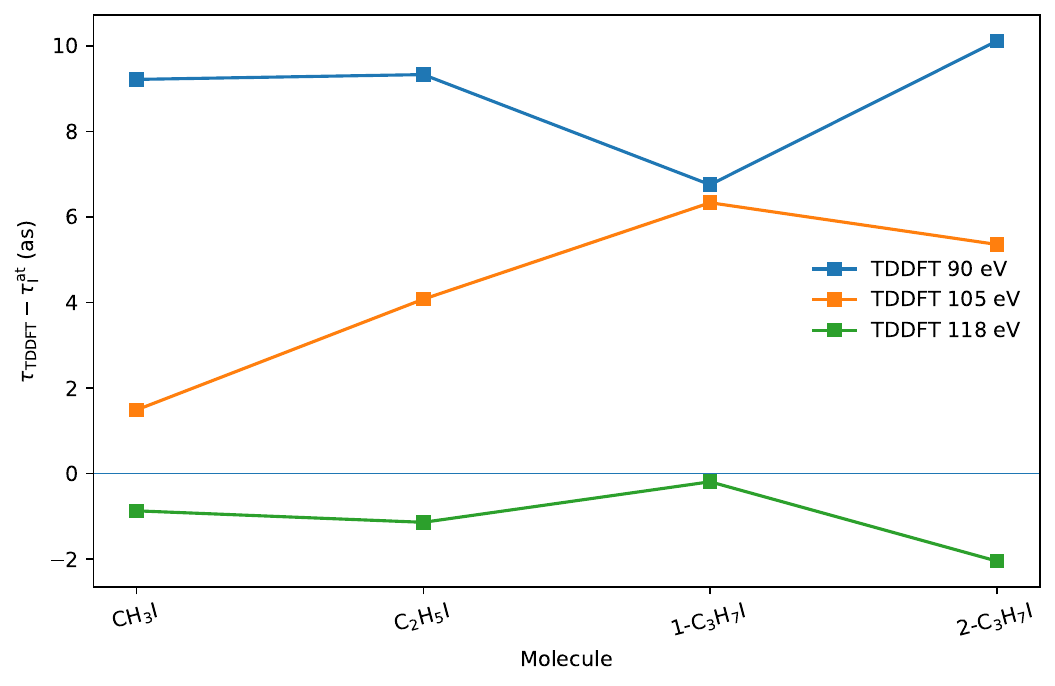}
\caption{Calculated field-free TDDFT departure from the atomic-I TDLDA
delay.  In contrast with \Fref{fig:expminus}, the theoretical
molecular dependence is small at all three photon energies.}
\label{fig:theoryminus}
\end{figure}

The contrast can be seen most clearly by subtracting the atomic reference.
Figure~\ref{fig:expminus} shows the experimental molecular shifts.  They
span approximately $-16$ to $+47$ as and change ordering with photon
energy.  Figure~\ref{fig:theoryminus} shows the same operation for TDDFT.
The entire theoretical set remains within roughly $-2$ to $+10$ as.

The most discriminating comparison is between 1- and 2-iodopropane.
At 90~eV the experiment gives about 37 and 79 as, respectively, while
TDDFT gives 39 and 42 as.  At 105~eV the experiment gives about 23 and
64 as, whereas TDDFT gives 36 and 35 as.  Thus the calculation not
only underestimates the experimental separation; it predicts
essentially no isomer splitting at all.  At 118~eV the theoretical and
experimental delays of the two propyl isomers move closer together,
although the absolute value for 1-iodopropane still differs
appreciably.

\begin{figure*}[t]
\centering
\includegraphics[width=0.75\textwidth]{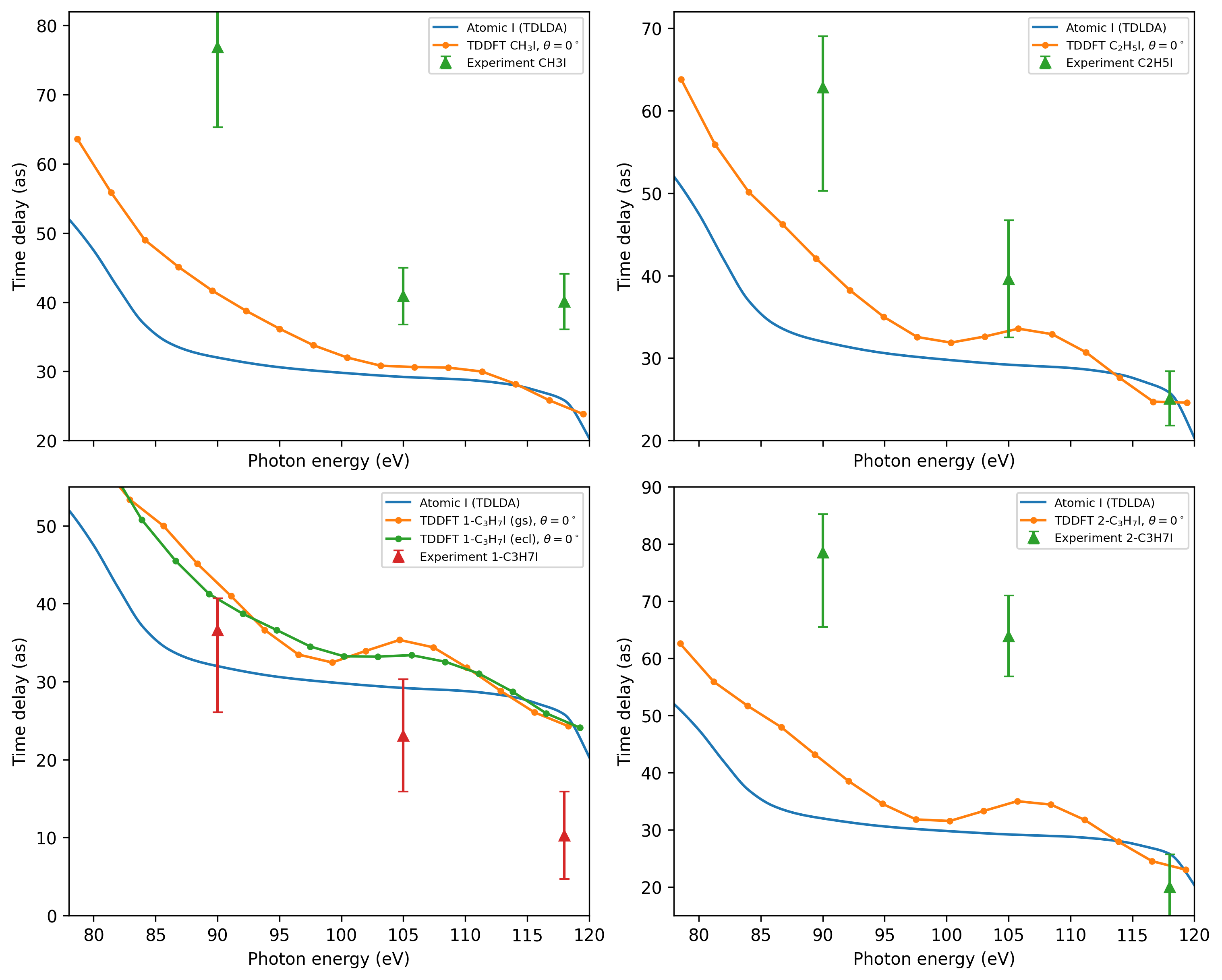}
\caption{Detailed comparison of the atomic-I TDLDA reference,
corrected TDDFT delay, and experimental streaking-derived delay for
CH$_3$I (upper left), C$_2$H$_5$I (upper right), 1-iodopropane
(lower left), and 2-iodopropane (lower right). For 1-iodopropane,
the ground-state and eclipsed conformers are shown separately, while
the experiment provides a single unresolved molecular value. The
corrected TDDFT curves are shown continuously over the 80--120~eV
range, whereas the experimental delays are available at 90, 105, and
118~eV.}
\label{fig:details}
\end{figure*}

For completeness, \Fref{fig:details} shows the detailed
molecule-by-molecule comparison underlying Table~\ref{tab:comparison}.
The corrected TDDFT curves are plotted continuously over the 80--120~eV
range, while the experimental streaking-derived delays remain available
at 90, 105, and 118~eV.  In each case the corrected molecular curve is
naturally compared with the atomic reference rather than with the raw,
convention-dependent delay.

\section{Discussion}
\label{sec:discussion}

\subsection{Resilience of the iodine $4d$ giant resonance}

Taken together, the atomic benchmarks and the molecular TDDFT calculations
suggest a simple physical picture.  Away from threshold, the broad iodine
$4d$ giant resonance is remarkably resilient.  Neutral atomic I and I$^-$
have very different asymptotic potentials, yet their calculated $4d$ delays
cluster together in the resonance region.  Attaching methyl, ethyl, or
propyl groups changes the molecular potential and introduces additional
scattering, but the corrected EWS delay remains close to the same atomic
scale.

This robustness is not accidental.  The corresponding cross sections and
$\beta$ parameters exhibit the same weak molecular sensitivity, and the
iodine $4d$ orbitals remain localized and nonbonding.  The dominant
$d\rightarrow f$ continuum therefore carries an iodine-centered collective
phase evolution whose energy scale is set primarily by the giant resonance,
not by the detailed alkyl geometry.  The field-free TDDFT ``bundling'' is
accordingly a physical result rather than a numerical failure.

A complementary indication follows from the logarithmic
Hilbert-transform relation between the modulus and phase of a
photoionization amplitude
\cite{PhysRevA.107.L021102,Ji2024,Kheifets2025resonant}.  Time delays
reconstructed from the energy dependence of the corresponding cross
sections do not necessarily agree quantitatively with directly
calculated Wigner delays, but their characteristic magnitude and
spectral variation are generally similar. The relation cannot be
applied rigorously to the total molecular cross section, where several
channels, partial waves, and interference terms contribute and
non-minimum-phase structure may be present.  Nevertheless, it provides
useful physical guidance: the rather weak molecular variation of the
giant-resonance cross section gives little indication of the
several-tens-of-attoseconds field-free delay variation required to
reproduce the experimental molecular dependence.

The present result also differs qualitatively from the recent
single-scattering model of \citet{Yao2026}. In that
description, the iodine-centered $4d\to\epsilon f$ wave acts as a
reference wave and interferes with components elastically scattered
from the carbon framework, producing a sizeable field-free difference
between the two iodopropane isomers. Structural scattering of this
kind is not absent from the present calculation. The \textsc{Tiresia}
continuum is generated in the full multicenter molecular potential and
therefore contains scattering from the complete molecular framework,
while the TDDFT response additionally includes self-consistent
screening and interchannel coupling. The fact that the resulting EWS
delays remain closely bundled therefore indicates that a
many-tens-of-attoseconds isomer splitting is not a robust consequence
of structural scattering alone.

The transparency of the single-scattering model comes at the price of
substantial simplification. It emphasizes the dominant
$4d\to\epsilon f$ channel and represents the subsequent propagation
through scattering from the molecular framework in a reduced form,
whereas the present calculation retains the full multicenter continuum
and its coupled molecular response. Such differences are particularly
important for a time delay, which is determined by an energy
derivative
of the photoionization phase and can therefore be sensitive to weak
amplitudes and interference terms that have only a modest influence on
the cross section.

There is also a potentially important difference in the treatment of
orientational averaging. In the present formulation the randomly
oriented delay is obtained from the orientationally averaged bilinear
quantities $\mathrm{Im}[A^*\partial_E A]$ and $|A|^2$, so that the
contribution of each molecular orientation is weighted by its
photoionization probability. This need not be equivalent to averaging
fixed-orientation phase derivatives themselves, particularly where
the photoionization amplitude becomes small and its phase varies
rapidly. The qualitative agreement of the reduced single-scattering
model with the measured isomer contrast therefore does not by itself
establish structural scattering in the field-free XUV step as the
origin of that contrast.

\subsection{What does the streaking measurement contain?}

For a simple Coulombic target the conventional streaking decomposition is
\begin{equation}
 \tau_{\rm st}=\tau_{\rm EWS}+\tau_{\rm CLC},
 \label{eq:conventional}
\end{equation}
where the Coulomb--laser-coupling (CLC), or continuum--continuum (CC), term
accounts for the universal long-range action of the IR field on the escaping
photoelectron \cite{Pazourek2015,Dahlstrom2012}.  \citet{Schroder2025} subtract this contribution to infer an ``absolute'' iodine $4d$ delay.  In the present molecular sequence the iodine $4d$
ionization potentials differ only weakly and every channel leaves a singly
charged residual ion.  The \emph{asymptotic Coulomb part} of the probe
correction is therefore expected to be nearly common to the molecules at a
fixed photon energy.  It is essential for the absolute calibration and is
not, by itself, a natural mechanism for the observed isomer-to-isomer
variation.  Equation~(\ref{eq:conventional}) should not, however, be read as
a general identity for a polyatomic target: molecular IR coupling can add
probe phases that are not contained in the universal Coulomb term.

A more cautious bookkeeping for the retrieved quantity is therefore
\begin{equation}
 \tau_{\rm st}^{\rm retrieved}
 =\tau_{\rm EWS}
 +\tau_{\rm CLC}^{\rm asym}
 +\delta\tau_{\rm IR}^{\rm mol}
 +\delta\tau_{\rm other},
 \label{eq:expandedstreaking}
\end{equation}
where $\delta\tau_{\rm IR}^{\rm mol}$ denotes target-specific IR-probe
dynamics beyond the asymptotic Coulomb correction, and
$\delta\tau_{\rm other}$ collects decay, correlation, spectral averaging,
and retrieval effects.  Neither term is separately measured in the present
experiment.

For comparison with the present calculation it is useful to define a residual
\begin{equation}
 \Delta\tau_{\rm residual}
 =\tau_{\rm st}^{\rm retrieved}
 -\tau_{\rm EWS}^{\rm TDDFT}-\tau_{\rm CLC}^{\rm asym}.
 \label{eq:general}
\end{equation}
Here $\tau_{\rm st}^{\rm retrieved}$ denotes the calibrated streaking delay
before subtraction of the asymptotic CLC contribution.  The experimental
``absolute'' delays tabulated in this work already include that subtraction;
for those values the same residual is
\begin{equation}
 \Delta\tau_{\rm residual}
 =\tau_{\rm exp}^{\rm abs}-\tau_{\rm EWS}^{\rm TDDFT}.
 \label{eq:residual_absolute}
\end{equation}
The CLC correction must therefore not be subtracted a second time.
This is a diagnostic difference, not an independently measured molecular-IR,
Auger, or PCI delay.  
%
A controlled synthetic-data test using the CH$_3$I TDDFT amplitudes
and a central-momentum Volkov probe model yields Gaussian-retrieval
delays within $0.5$~as of the bandwidth- and acceptance-averaged input
delays; allowing joint reconstruction of an idealized reference trace
and the IR field at $105$~eV leaves this agreement essentially
unchanged. This internal validation does
not include molecular CC/CLC effects or reproduce the experimental
spin--orbit structure and noise.
The present calculations do not separate the
contributions on the right-hand side of Eq.~(\ref{eq:expandedstreaking}).
Agreement in $\sigma$ and $\beta$ is an important consistency check but does
not by itself validate either the field-free spectral phase or the full
probe response, as the common-phase example in Sec.~\ref{sec:slb} and the
molecular two-color results discussed below both illustrate.

\subsection{Molecular IR-probe effects beyond asymptotic CLC}
\label{sec:molecular-ir}

Two recent calculations give a useful warning against identifying the
complete molecular probe correction with $\tau_{\rm CLC}^{\rm asym}$.
\citet{Delgado2025} solved the multielectron TDSE for RABBITT in
acetylene and found an additional pathway in which the XUV photon emits the
photoelectron while the IR photon drives a transition between two cationic
states.  This third path changes the extracted two-photon
delay and makes it sensitive to the IR frequency.  Crucially for the present
comparison, the effect is strongly orientation selective: it is large for
the molecular axis parallel to the polarization, absent for the
perpendicular component in the channels considered, and consequently much
smaller, although still visible, for randomly oriented molecules.

\citet{Pranjal2026} reach a complementary conclusion from fully
correlated calculations of H$_2$O.  There the relevant
IR effect is not a simple isotropic displacement of the Wigner delay.
Continuum--continuum coupling reshapes the angular structure, producing
molecular-frame variations as large as about 80 as between opposite emission
directions.  They also show that similar photoelectron angular distributions
do not guarantee similar time-delay structures.  This point is directly
relevant to the present observation that the iodoalkane cross sections and
$\beta$ parameters are very similar: conventional one-photon observables do
not by themselves exclude a phase-sensitive molecular IR correction.

There are, however, equally important reasons not to transfer these large
molecular-frame values directly to the iodoalkane streaking data.  Both
studies concern RABBIT rather than streaking; the acetylene effect is reduced
by random-orientation averaging, and the H$_2$O work explicitly notes that
orientation averaging can partially wash out its intrinsic angular
structure.  Moreover, the latter calculations concern substantially lower
photoelectron energies and anticipate weaker molecular-continuum anisotropy
for faster electrons.  The iodine $4d$ photoelectrons in the present
experiment have kinetic energies of roughly 31--59~eV at the three photon
energies, so a many-tens-of-attoseconds molecular probe correction in the
orientation-averaged streaking signal is possible in principle but is not
established by these examples.

A distinct possible contribution is the near field generated by IR-induced
polarization of the residual molecular ion \cite{Pazourek2015}.  Unlike an
unweighted permanent-dipole vector, the scalar part of the polarizability
survives random molecular orientation, where $R$ denotes the rotation from
the molecular to the laboratory frame:
\begin{equation}
 \begin{aligned}
 \left\langle R\boldsymbol{\alpha}^{+}(\omega)R^{\mathsf T}\right\rangle_R
 &=\bar\alpha^{+}(\omega)\mathbf{1},\\
 \bar\alpha^{+}&=\tfrac13\operatorname{Tr}\boldsymbol{\alpha}^{+}.
 \end{aligned}
 \label{eq:ir_isotropic_response}
\end{equation}
The brackets denote a uniform orientation average.  The induced dipole
produces an exterior electron--ion
interaction $\delta V({\bf r},t)=-{\bf p}(t)\cdot{\bf r}/r^3$, distinct
from the electron-induced polarization potential proportional to $r^{-4}$.
This IR-driven response is not determined by the present one-photon TDDFT
amplitudes.

Its survival under orientation averaging does not establish a large
streaking delay.  The electron samples both the inner induced field and its
exterior dipole tail, whose effects can partially cancel.  The neutral--ion
differential dressing and the transient response following creation of the
$4d$ vacancy must also be treated consistently.  Consequently, neither a
large total polarizability nor an exterior dipole-force estimate alone
determines the magnitude or sign of the timing correction.  Such a
correction could reflect a change in the streaking phase without additional
electron trapping.

IR-driven ionic transitions provide another possible contribution
\cite{BendaMasin2024,Makos2025}, but their effect depends on the measurement
projection.  The CO$_2$ experiment of \citet{Makos2025} selected a final
ionic channel through coincidence detection.  By contrast, a unitary
evolution acting only on an already decoupled ion leaves the fully
ion-summed electron spectrum unchanged.  Application of this mechanism to
the present experiment therefore requires an explicit treatment of the
ionization pathways, electron--ion interaction, and unobserved ionic outcomes.

The decisive test remains a calculation of XUV+IR iodoalkane spectrograms,
summed over molecular orientations and unresolved ionic channels and
integrated over the experimental angular acceptance before applying the
streaking retrieval.  Controlled comparisons should distinguish
induced-density response, explicit ionic transitions, and post-ionization
Coulomb dynamics without double counting.  Until such calculations reproduce
the magnitude, sign, and isomer dependence of $\Delta\tau_{\rm residual}$,
molecular IR dynamics remain a candidate contribution rather than an
established explanation.

\subsection{NVV Auger--Meitner decay and the contamination test}
\label{sec:auger}

Recent attosecond X-ray measurements of core-level photoionization in NO
have demonstrated substantial contributions from molecular shape resonances,
interchannel coupling, and post-collision interaction with Auger--Meitner
electrons \cite{Driver2024}.

The relevant sequence is
\begin{equation}
\begin{aligned}
 ({\rm R-I})+\hbar\omega
 &\longrightarrow ({\rm R-I})^+(4d^{-1})+e_{4d}\\
 &\longrightarrow ({\rm R-I})^{2+}(V^{-2})+e_{4d}+e_{\rm A}.
\end{aligned}
 \label{eq:augersequence}
\end{equation}
The XUV pulse creates the core hole; subsequent Auger decay is driven by
the electron--electron interaction.  The energies and branching ratios of
the valence-hole final states can depend on the ligand.  In a spectator
approximation, however, this decay need not substantially alter the primary
photoionization cross section or angular distribution.  Its presence alone
therefore does not imply a correction to the primary streaking delay.

\citet{Schroder2025} simulated streaked NVV emission overlapping the
I $4d$ signal at 90~eV, varied decay lifetimes and spectral structure, and
applied their retrieval to the summed spectrogram.  Their maximum bias was
6.33 as for a 1.5 fs lifetime, included as an additional downward systematic
uncertainty.  This is a substantive test of additive
Auger contamination, not merely a subtraction of a static background.

To see what this procedure includes, let $Q_j(E,\theta)$ be the complex
fundamental-frequency component of the primary ($j={\rm P}$) or Auger
($j={\rm A}$) spectrogram, and let $D_j$ denote the corresponding
field-free emission amplitude.  At lowest order in the IR field its
interference structure is schematically \cite{Dahlstrom2012}
\begin{equation}
 Q_j=D_jM_j^{+*}+D_j^*M_j^-,
 \label{eq:heterodyne}
\end{equation}
where $M_j^\pm$ denote the IR-absorption and IR-emission companions.
For Auger emission this notation suppresses sums over ionic channels and
unobserved-electron variables.  The one-IR interference is inversion odd
for an isotropic ensemble in the electric-dipole approximation.  Both
primary and Auger emission can therefore contribute to the same odd signal.
Nevertheless, adding their intensities already gives
\begin{equation}
 S=S_{\rm P}+S_{\rm A}
 \quad\Longrightarrow\quad Q=Q_{\rm P}+Q_{\rm A}.
 \label{eq:phasoraddition}
\end{equation}
Their modulations can change the measured phase without coherent
interference between primary and Auger electron amplitudes.  The delay
derivative used in the experimental fit preserves this addition.
Consequently, Eq.~(\ref{eq:phasoraddition}) does not reveal a missing term
in the published contamination test.

The magnitude of a bias depends on the complex $Q_{\rm A}$ in the fitted
energy and angular window, not just the integrated Auger yield.  For a
local single-frequency observable, writing
$Q_{\rm A}/Q_{\rm P}=\rho\exp(i\phi)$ gives
\begin{equation}
 \delta\tau=\frac{1}{\omega_{\rm IR}}
 \arg\left[1+\rho\exp(i\phi)\right],
 \label{eq:localbias}
\end{equation}
with a convention in which positive phase denotes positive delay.
Equation~(\ref{eq:localbias}) is not a replacement for a full spectrogram
fit.  The reported 6.33 as value is conditional on the simulated Auger
response and retrieval; the present work neither reproduces nor overturns
that quantitative estimate.

The described additive test does not establish a bound on coupled
post-ionization Coulomb dynamics.  This limitation is not positive evidence
for PCI.  Any proposed correction must reproduce the actual spectrogram
and its molecular dependence, including photon energies at which the
primary and dominant Auger bands are separated.

\subsection{PCI: velocity ordering and emission separation}
\label{sec:pci}

Field-free PCI theory establishes energy redistribution and line-shape
changes following core-hole decay
\cite{Kuchiev1985,Kuchiev1986CPC,Kuchiev1988,Kuchiev1989}.  In the familiar
overtaking regime a faster Auger electron passes the slower primary
electron, changing the screening of the ionic charge.  The primary loses
energy while the Auger electron gains it.  Small-relative-velocity Coulomb
parameters can signal strong kinematic sensitivity, but do not alone
specify the encounter distance or the observable streaking-phase shift.
In particular, a formal $1/v_{\rm rel}$ scaling cannot be used as a
quantitative amplification factor for sequential emission without checking
its assumptions and the finite separation at Auger birth.

Taking a representative iodine $4d$ binding energy of 58.6~eV and an NVV
Auger energy of 30~eV gives
\begin{center}
\begin{tabular}{ccccc}
\hline\hline
$\hbar\omega$ & $E_{4d}$ & $v_{4d}$ & $v_{\rm A}$ &
$v_{4d}-v_{\rm A}$\\
(eV) & (eV) & (a.u.) & (a.u.) & (a.u.)\\
\hline
90  & 31.4 & 1.519 & 1.485 & 0.034\\
105 & 46.4 & 1.847 & 1.485 & 0.362\\
118 & 59.4 & 2.089 & 1.485 & 0.605\\
\hline\hline
\end{tabular}
\end{center}
The representative primary electron is faster at all three energies.
The lower-energy part of the finite-bandwidth primary spectrum at 90~eV
can permit overtaking, but conventional catch-up is less favorable for the
central primary bands at 105 and 118~eV.  Near equality of speeds at 90~eV
must therefore not be identified with an established PCI enhancement.

A simple radial, constant-speed estimate illustrates the role of emission
separation.  If the Auger electron is emitted a time $s$ after the primary,
then, for $v_{\rm A}>v_{4d}$,
\begin{equation}
 t_c-t_{\rm A}=\frac{v_{4d}\, s}{v_{\rm A}-v_{4d}},
 \qquad
 r_c=\frac{v_{\rm A}v_{4d}\, s}{v_{\rm A}-v_{4d}}.
 \label{eq:catchup}
\end{equation}
The characteristic screening energy transfer scales as $1/r_c$ in atomic
units.  For fixed $s$, near equality of speeds moves overtaking farther
from the ion rather than guaranteeing a larger energy transfer.  As an
illustration, a 28~eV primary electron, a 30~eV Auger electron, and
$s=1.5$ fs give $t_c-t_{\rm A}\simeq43$ fs and
$r_c\simeq2.6\times10^3 a_0$, corresponding to about 0.01~eV.
This estimate neglects Coulomb acceleration, the IR field, and angular
correlations and is not an upper bound: earlier decays and other
trajectories can give stronger effects.  It also does not exclude PCI
outside the overtaking approximation.  It does show why matching the
asymptotic speeds is insufficient to infer a large delay.

\subsection{Field-assisted PCI and the retrieved phase}
\label{sec:pci-streaking}

The IR field can modify the relative motion because the primary electron is
emitted first.  Neglecting Coulomb forces temporarily and using
$\bm E=-\partial_t\bm A$, its momentum at Auger birth is
\begin{equation}
 \bm p_{4d}(t_{\rm A})\simeq\bm p_{4d}^{(0)}
 +[\bm A(t_{\rm A})-\bm A(t_0)].
\end{equation}
For parallel emission the initial relative velocity is therefore
\begin{equation}
 v_{\rm A}-v_{4d}^{(0)}-A(t_{\rm A})-A(t_0).
 \label{eq:ircatchup}
\end{equation}
A favorable vector-potential difference can reverse the field-free speed
ordering.  In this simplified estimate the required differences for the
central energies above are 0.034, 0.362, and 0.605 a.u., respectively.
Whether these conditions occur with significant probability must be checked
against the reconstructed pulse and decay-time distribution.  A spatially
homogeneous field exerts equal forces on both electrons once both are free;
their subsequent relative acceleration arises from their Coulomb forces.
The primary electron's spatial head start must also be included.

\citet{BauchBonitz2012} showed that field-assisted PCI can modify Auger
streaking line shapes, but found primary-electron broadening without the
corresponding Auger-like chirp under the conditions they studied.  That result is not a universal exclusion for the present
molecules, but demonstrates that an Auger distortion cannot automatically
be interpreted as a primary-photoelectron timing shift.

A delay-independent energy displacement can be absorbed into a fitted line
position.  Establishing a timing bias instead requires evaluating how the
PCI-induced changes project onto the delay-sensitive part of the actual
retrieval.  Although a 30 as shift at 790 nm corresponds to only about
0.07 rad, this conversion says nothing about whether the required
phase-dependent distortion is generated.  No such calculation has been
performed here.  In particular, the absence of PCI from an additive model
does not establish that it can explain the isomer dependence at 105~eV or
the discrepancies at 118~eV.

\subsection{Scope of the Xe/Kr comparison}
\label{sec:hasan}

\citet{Hasan2026} report an anomalous Xe Auger streaking shift of about
1.32 fs relative to the $4d$ photoelectrons, together with a minimum of the
Auger yield near a pump--probe delay of 4 fs.  These
observations motivate scrutiny of Auger streaking models, but they do not
identify PCI as their cause or demonstrate a corresponding correction to
an iodine primary-photoelectron delay.  The reported absence of analogous
anomalies in Kr is an additional reason not to infer a universal mechanism
from similar electron speeds.  The atomic experiments provide useful
benchmarks for future coupled calculations, not evidence that Auger/PCI
explains the molecular ordering in the present comparison.

\subsection{A conditional test of Auger/PCI}
\label{sec:pci-tests}

A useful test would compare otherwise identical XUV+IR calculations with
and without post-ionization Coulomb coupling.  The primary electron would
be launched with the calculated photoionization distribution, the Auger
emission time sampled from the decay law, and both electrons propagated
with the residual ion and the reconstructed IR field.  An initial
semiclassical model could screen the importance of the effect, while its
short-range and quantum limitations would need separate assessment.

The detector response should include both electrons, the experimental
energy windows and angular acceptance.  Applying the same retrieval to
these synthetic spectrograms defines
\begin{equation}
 \Delta\tau_{\rm PCI}
 =\tau_{\rm retrieved}^{\rm PCI}
 -\tau_{\rm retrieved}^{\rm no\,PCI}.
\end{equation}
Separately inspecting the primary and Auger contributions would distinguish
changes of primary propagation from changes of additive contamination.
Molecule-specific Auger energies, branching ratios, lifetimes, and angular
correlations would then be required to test chemical dependence.
The comparison must address all three photon energies, rather than only
the near-overlap region at 90~eV.  It should also remain compatible with
the field-free spectra.  Until such a test demonstrates the required
magnitude, sign, and molecular ordering, Auger/PCI remains an unconfirmed
possibility among the mechanisms and modeling limitations that could
contribute to the discrepancy.

\section{Conclusions}
\label{sec:conclusions}

We have developed and tested a direct evaluation of the molecular
Eisenbud--Wigner--Smith time delay from complex B-spline TDDFT
photoionization amplitudes in the iodine $4d$ giant-resonance region.  The
delay is expressed through rank-0 and rank-2 tensor contractions, allowing
the same angular algebra used for the cross section and anisotropy parameter
to be reused for the phase derivative.

The phase sensitivity of the observable exposed an asymptotic convention in
the LB94 continuum matching.  The shifted Coulomb coordinate introduces a
smooth common energy-dependent phase into the $S$-matrix-normalized dipoles.
Because this factor has unit modulus, it is completely invisible in
$\sigma$ and $\beta$ but gives a large spurious background to
$\partial_E\arg D$.  Reinstating the corresponding phase convention removes
this background and returns the molecular delays to the expected atomic
scale.  The correction is fixed by the continuum matching and is not fitted
to the streaking experiment.

The corrected field-free result is robust: CH$_3$I, C$_2$H$_5$I,
2-iodopropane, and both the ground-state and eclipsed conformers of
1-iodopropane produce closely clustered EWS delays throughout the
90--118~eV experimental range.  Neutral atomic I, I$^-$, RPAE/RRPA,
and molecular TDDFT show the same broad giant-resonance phase trend
away from threshold.  We therefore conclude that the iodine $4d$ giant
resonance is remarkably resilient to both charge state and chemical
environment.  Molecular corrections exist, but they are much smaller
than the large, nonmonotonic and isomer-dependent variations reported
by attosecond streaking.

The present results also do not support the recent field-free
single-scattering interpretation of \citet{Yao2026}. Their
model attributes a sizeable iodopropane isomer contrast to
interference between the iodine-centered outgoing wave and components
scattered from the carbon framework. In the present calculation such
structural scattering is already contained in the full multicenter
molecular continuum, together with self-consistent screening and
interchannel coupling, yet no comparable isomer splitting
emerges. This conclusion is independently consistent with the weak
molecular variation of the giant-resonance cross sections and with the
characteristic delay scale suggested by their modulus--phase
relation. We therefore find no evidence that the observed
several-tens-of-attoseconds isomer contrast is a robust consequence of
field-free structural scattering alone.  Qualitative agreement of a
reduced single-scattering model with the experimental contrast does
not, by itself, establish that mechanism as its physical origin.

The mismatch is unlikely to be generated by the \emph{standard
  asymptotic} CLC/CC correction.  At a fixed photon energy that
contribution is controlled primarily by the outgoing electron energy
and residual charge, both of which are nearly the same throughout the
iodoalkane sequence.  Target-specific molecular IR dynamics
nevertheless remain a physically motivated missing contribution,
subject to the qualifications discussed in
Sec.~\ref{sec:molecular-ir}.

The origin of the remaining molecular variation is therefore
unresolved.  NVV Auger--Meitner decay provides another possible
connection to valence-shell chemistry, but neither its presence nor
its molecular sensitivity establishes a correction to the
primary-photoelectron delay.  The published Auger contamination test
already includes addition of the primary and Auger streaking
modulations and finds a maximum bias of 6.33 as within its tested
model \cite{Schroder2025}.  We identify no omitted interference term
that would invalidate that procedure.

Coupled PCI dynamics is not bounded by that additive test, but its
importance cannot be inferred from near velocity matching alone.  The
central primary electrons are faster than a representative 30~eV Auger
electron at all three photon energies, with increasing separation at 105
and 118~eV.  Even when overtaking is possible, the emission-time separation
can place the encounter far from the ion.  The IR may modify these
conditions, but the resulting delay bias requires a calculation with the
actual pulse, angular acceptance, and retrieval.  The anomalous Xe Auger
streaking observations \cite{Hasan2026} do not by themselves establish PCI
as the missing mechanism in the iodoalkanes.

The present work therefore establishes a field-free reference and a
quantitative theory--experiment discrepancy, rather than a
demonstrated molecular-IR or Auger/PCI explanation.  The most direct
next step is an orientation-averaged XUV+IR streaking calculation for
the same iodoalkanes, with the experimental pulse and retrieval,
followed by controlled variants that separately suppress the
IR-induced density response, explicit ionic transitions, and
post-ionization electron--electron coupling, while avoiding double
counting between these descriptions.  Such a comparison would separate
target-specific probe dynamics from decay effects and from limitations
of the one-photon treatment and retrieval.  The broader need for an
explicit two-color description is further underscored by recent
RABBITT measurements on CH$_3$I, providing an interferometric probe of
photoionization dynamics in this prototypical iodine-containing
molecule \cite{Carneiro2026}.

\begin{acknowledgments}
We thank Maximilian Forster and Reinhard Kienberger for sharing
their experimental results prior to publication and for helpful
discussions. Computational resources were provided by the National
Computational Infrastructure (NCI) through the ANU Merit Allocation
Scheme. General support for travel and scientific exchange was
provided by the Australian Research Council through Discovery
Project DP230101253.
\end{acknowledgments}

\section*{Data Availability}

The numerical TDDFT data supporting this study, including
orbital- and symmetry-resolved photoionization cross sections,
anisotropy parameters, and EWS/Wigner time delays for the
investigated iodoalkanes, are publicly available in the
Zenodo repository \cite{Kheifets2026Zenodo}.


\begin{thebibliography}{62}
\expandafter\ifx\csname natexlab\endcsname\relax\def\natexlab#1{#1}\fi
\expandafter\ifx\csname bibnamefont\endcsname\relax
  \def\bibnamefont#1{#1}\fi
\expandafter\ifx\csname bibfnamefont\endcsname\relax
  \def\bibfnamefont#1{#1}\fi
\expandafter\ifx\csname citenamefont\endcsname\relax
  \def\citenamefont#1{#1}\fi
\expandafter\ifx\csname url\endcsname\relax
  \def\url#1{\texttt{#1}}\fi
\expandafter\ifx\csname urlprefix\endcsname\relax\def\urlprefix{URL }\fi
\providecommand{\bibinfo}[2]{#2}
\providecommand{\eprint}[2][]{\url{#2}}

\bibitem[{\citenamefont{Wigner}(1955)}]{Wigner1955}
\bibinfo{author}{\bibfnamefont{E.~P.} \bibnamefont{Wigner}},
  \emph{\bibinfo{title}{Lower limit for the energy derivative of the scattering
  phase shift}}, \bibinfo{journal}{Phys. Rev.} \textbf{\bibinfo{volume}{98}},
  \bibinfo{pages}{145} (\bibinfo{year}{1955}).

\bibitem[{\citenamefont{Pazourek et~al.}(2015)\citenamefont{Pazourek, Nagele,
  and Burgdorfer}}]{Pazourek2015}
\bibinfo{author}{\bibfnamefont{R.}~\bibnamefont{Pazourek}},
  \bibinfo{author}{\bibfnamefont{S.}~\bibnamefont{Nagele}}, \bibnamefont{and}
  \bibinfo{author}{\bibfnamefont{J.}~\bibnamefont{Burgdorfer}},
  \emph{\bibinfo{title}{Attosecond chronoscopy of photoemission}},
  \bibinfo{journal}{Rev. Mod. Phys.} \textbf{\bibinfo{volume}{87}},
  \bibinfo{pages}{765} (\bibinfo{year}{2015}).

\bibitem[{\citenamefont{Kheifets}(2023)}]{Kheifets2023review}
\bibinfo{author}{\bibfnamefont{A.~S.} \bibnamefont{Kheifets}},
  \emph{\bibinfo{title}{Wigner time delay in atomic photoionization}},
  \bibinfo{journal}{J.~Phys.~B}
  \textbf{\bibinfo{volume}{56}}(\bibinfo{number}{2}), \bibinfo{pages}{022001}
  (\bibinfo{year}{2023}).

\bibitem[{\citenamefont{Itatani et~al.}(2002)\citenamefont{Itatani, Quere,
  Yudin, Ivanov, Krausz, and Corkum}}]{Itatani2002}
\bibinfo{author}{\bibfnamefont{J.}~\bibnamefont{Itatani}},
  \bibinfo{author}{\bibfnamefont{F.}~\bibnamefont{Quere}},
  \bibinfo{author}{\bibfnamefont{G.~L.} \bibnamefont{Yudin}},
  \bibinfo{author}{\bibfnamefont{M.~Y.} \bibnamefont{Ivanov}},
  \bibinfo{author}{\bibfnamefont{F.}~\bibnamefont{Krausz}}, \bibnamefont{and}
  \bibinfo{author}{\bibfnamefont{P.~B.} \bibnamefont{Corkum}},
  \emph{\bibinfo{title}{Attosecond streak camera}}, \bibinfo{journal}{Phys.
  Rev. Lett.} \textbf{\bibinfo{volume}{88}}, \bibinfo{pages}{173903}
  (\bibinfo{year}{2002}).

\bibitem[{\citenamefont{Yakovlev et~al.}(2010)\citenamefont{Yakovlev, Gagnon,
  Karpowicz, and Krausz}}]{Yakovlev2010}
\bibinfo{author}{\bibfnamefont{V.~S.} \bibnamefont{Yakovlev}},
  \bibinfo{author}{\bibfnamefont{J.}~\bibnamefont{Gagnon}},
  \bibinfo{author}{\bibfnamefont{N.}~\bibnamefont{Karpowicz}},
  \bibnamefont{and} \bibinfo{author}{\bibfnamefont{F.}~\bibnamefont{Krausz}},
  \emph{\bibinfo{title}{Attosecond streaking enables the measurement of quantum
  phase}}, \bibinfo{journal}{Phys. Rev. Lett.} \textbf{\bibinfo{volume}{105}},
  \bibinfo{pages}{073001} (\bibinfo{year}{2010}).

\bibitem[{\citenamefont{Dahlstrom et~al.}(2013)\citenamefont{Dahlstrom, Guenot,
  Klunder, Gisselbrecht, Mauritsson, L'Huillier, Maquet, and
  Taieb}}]{Dahlstrom2012}
\bibinfo{author}{\bibfnamefont{J.~M.} \bibnamefont{Dahlstrom}},
  \bibinfo{author}{\bibfnamefont{D.}~\bibnamefont{Guenot}},
  \bibinfo{author}{\bibfnamefont{K.}~\bibnamefont{Klunder}},
  \bibinfo{author}{\bibfnamefont{M.}~\bibnamefont{Gisselbrecht}},
  \bibinfo{author}{\bibfnamefont{J.}~\bibnamefont{Mauritsson}},
  \bibinfo{author}{\bibfnamefont{A.}~\bibnamefont{L'Huillier}},
  \bibinfo{author}{\bibfnamefont{A.}~\bibnamefont{Maquet}}, \bibnamefont{and}
  \bibinfo{author}{\bibfnamefont{R.}~\bibnamefont{Taieb}},
  \emph{\bibinfo{title}{Theory of attosecond delays in laser-assisted
  photoionization}}, \bibinfo{journal}{Chem. Phys.}
  \textbf{\bibinfo{volume}{414}}, \bibinfo{pages}{53} (\bibinfo{year}{2013}).

\bibitem[{\citenamefont{Nahon et~al.}(1991)\citenamefont{Nahon, Svensson, and
  Morin}}]{Nahon1991}
\bibinfo{author}{\bibfnamefont{L.}~\bibnamefont{Nahon}},
  \bibinfo{author}{\bibfnamefont{A.}~\bibnamefont{Svensson}}, \bibnamefont{and}
  \bibinfo{author}{\bibfnamefont{P.}~\bibnamefont{Morin}},
  \emph{\bibinfo{title}{Experimental study of the 4d ionization continuum in
  atomic iodine by photoelectron and photoion spectroscopy}},
  \bibinfo{journal}{Phys. Rev. A} \textbf{\bibinfo{volume}{43}},
  \bibinfo{pages}{2328} (\bibinfo{year}{1991}).

\bibitem[{\citenamefont{O'Sullivan et~al.}(1996)\citenamefont{O'Sullivan,
  McGuinness, Costello, Kennedy, and Weinmann}}]{OSullivan1996}
\bibinfo{author}{\bibfnamefont{G.}~\bibnamefont{O'Sullivan}},
  \bibinfo{author}{\bibfnamefont{C.}~\bibnamefont{McGuinness}},
  \bibinfo{author}{\bibfnamefont{J.~T.} \bibnamefont{Costello}},
  \bibinfo{author}{\bibfnamefont{E.~T.} \bibnamefont{Kennedy}},
  \bibnamefont{and} \bibinfo{author}{\bibfnamefont{B.}~\bibnamefont{Weinmann}},
  \emph{\bibinfo{title}{Trends in 4d-subshell photoabsorption along the iodine
  isonuclear sequence: {I}, {I}$^+$, and {I}$^{2+}$}}, \bibinfo{journal}{Phys.
  Rev. A} \textbf{\bibinfo{volume}{53}}, \bibinfo{pages}{3211}
  (\bibinfo{year}{1996}).

\bibitem[{\citenamefont{Pi and Landsman}(2018)}]{PiLandsman2018}
\bibinfo{author}{\bibfnamefont{L.-W.} \bibnamefont{Pi}} \bibnamefont{and}
  \bibinfo{author}{\bibfnamefont{A.~S.} \bibnamefont{Landsman}},
  \emph{\bibinfo{title}{Attosecond time delay in photoionization of noble-gas
  and halogen atoms}}, \bibinfo{journal}{Appl. Sci.}
  \textbf{\bibinfo{volume}{8}}, \bibinfo{pages}{322} (\bibinfo{year}{2018}).

\bibitem[{\citenamefont{Magrakvelidze et~al.}(2016)\citenamefont{Magrakvelidze,
  Madjet, and Chakraborty}}]{Magrakvelidze2016}
\bibinfo{author}{\bibfnamefont{M.}~\bibnamefont{Magrakvelidze}},
  \bibinfo{author}{\bibfnamefont{M.~E.-A.} \bibnamefont{Madjet}},
  \bibnamefont{and} \bibinfo{author}{\bibfnamefont{H.~S.}
  \bibnamefont{Chakraborty}}, \emph{\bibinfo{title}{Attosecond delay of xenon
  4d photoionization at the giant resonance and cooper minimum}},
  \bibinfo{journal}{Phys. Rev. A} \textbf{\bibinfo{volume}{94}},
  \bibinfo{pages}{013429} (\bibinfo{year}{2016}).

\bibitem[{\citenamefont{Mandal et~al.}(2017)\citenamefont{Mandal, Deshmukh,
  Kheifets, Dolmatov, and Manson}}]{PhysRevA.96.053407}
\bibinfo{author}{\bibfnamefont{A.}~\bibnamefont{Mandal}},
  \bibinfo{author}{\bibfnamefont{P.~C.} \bibnamefont{Deshmukh}},
  \bibinfo{author}{\bibfnamefont{A.~S.} \bibnamefont{Kheifets}},
  \bibinfo{author}{\bibfnamefont{V.~K.} \bibnamefont{Dolmatov}},
  \bibnamefont{and} \bibinfo{author}{\bibfnamefont{S.~T.}
  \bibnamefont{Manson}}, \emph{\bibinfo{title}{Angle-resolved {Wigner} time
  delay in atomic photoionization: The $4d$ subshell of free and confined
  {Xe}}}, \bibinfo{journal}{Phys. Rev. A} \textbf{\bibinfo{volume}{96}},
  \bibinfo{pages}{053407} (\bibinfo{year}{2017}).

\bibitem[{\citenamefont{Bray et~al.}(2018)\citenamefont{Bray, Naseem, and
  Kheifets}}]{PhysRevA.98.043427}
\bibinfo{author}{\bibfnamefont{A.~W.} \bibnamefont{Bray}},
  \bibinfo{author}{\bibfnamefont{F.}~\bibnamefont{Naseem}}, \bibnamefont{and}
  \bibinfo{author}{\bibfnamefont{A.~S.} \bibnamefont{Kheifets}},
  \emph{\bibinfo{title}{Photoionization of {Xe} and {Xe}@{C}$_{60}$ from the
  $4d$ shell in {RABBITT} fields}}, \bibinfo{journal}{Phys. Rev. A}
  \textbf{\bibinfo{volume}{98}}, \bibinfo{pages}{043427}
  (\bibinfo{year}{2018}).

\bibitem[{\citenamefont{Ossiander et~al.}(2018)\citenamefont{Ossiander,
  Riemensberger, Neppl, Mittermair, Schaffer, Duensing, Wagner, Heider, Wurzer,
  Gerl et~al.}}]{Ossiander2018}
\bibinfo{author}{\bibfnamefont{M.}~\bibnamefont{Ossiander}},
  \bibinfo{author}{\bibfnamefont{J.}~\bibnamefont{Riemensberger}},
  \bibinfo{author}{\bibfnamefont{S.}~\bibnamefont{Neppl}},
  \bibinfo{author}{\bibfnamefont{M.}~\bibnamefont{Mittermair}},
  \bibinfo{author}{\bibfnamefont{M.}~\bibnamefont{Schaffer}},
  \bibinfo{author}{\bibfnamefont{A.}~\bibnamefont{Duensing}},
  \bibinfo{author}{\bibfnamefont{M.~S.} \bibnamefont{Wagner}},
  \bibinfo{author}{\bibfnamefont{R.}~\bibnamefont{Heider}},
  \bibinfo{author}{\bibfnamefont{M.}~\bibnamefont{Wurzer}},
  \bibinfo{author}{\bibfnamefont{M.}~\bibnamefont{Gerl}}, \bibnamefont{et~al.},
  \emph{\bibinfo{title}{Absolute timing of the photoelectric effect}},
  \bibinfo{journal}{Nature} \textbf{\bibinfo{volume}{561}},
  \bibinfo{pages}{374} (\bibinfo{year}{2018}).

\bibitem[{\citenamefont{Biswas et~al.}(2020)\citenamefont{Biswas, Forg,
  Ortmann, Schotz, Schweinberger, Zimmermann, Pi, Baykusheva, Masood, Liontos
  et~al.}}]{Biswas2020}
\bibinfo{author}{\bibfnamefont{S.}~\bibnamefont{Biswas}},
  \bibinfo{author}{\bibfnamefont{B.}~\bibnamefont{Forg}},
  \bibinfo{author}{\bibfnamefont{L.}~\bibnamefont{Ortmann}},
  \bibinfo{author}{\bibfnamefont{J.}~\bibnamefont{Schotz}},
  \bibinfo{author}{\bibfnamefont{W.}~\bibnamefont{Schweinberger}},
  \bibinfo{author}{\bibfnamefont{T.}~\bibnamefont{Zimmermann}},
  \bibinfo{author}{\bibfnamefont{L.-W.} \bibnamefont{Pi}},
  \bibinfo{author}{\bibfnamefont{D.}~\bibnamefont{Baykusheva}},
  \bibinfo{author}{\bibfnamefont{H.~A.} \bibnamefont{Masood}},
  \bibinfo{author}{\bibfnamefont{I.}~\bibnamefont{Liontos}},
  \bibnamefont{et~al.}, \emph{\bibinfo{title}{Probing molecular environment
  through photoemission delays}}, \bibinfo{journal}{Nat. Phys.}
  \textbf{\bibinfo{volume}{16}}, \bibinfo{pages}{778} (\bibinfo{year}{2020}).

\bibitem[{\citenamefont{Schr\"oder et~al.}(2025)\citenamefont{Schr\"oder,
  Pollanka, Freisinger, Ostner, Forster, Paul, and Kienberger}}]{Schroder2025}
\bibinfo{author}{\bibfnamefont{C.~A.} \bibnamefont{Schr\"oder}},
  \bibinfo{author}{\bibfnamefont{M.}~\bibnamefont{Pollanka}},
  \bibinfo{author}{\bibfnamefont{P.}~\bibnamefont{Freisinger}},
  \bibinfo{author}{\bibfnamefont{M.}~\bibnamefont{Ostner}},
  \bibinfo{author}{\bibfnamefont{M.}~\bibnamefont{Forster}},
  \bibinfo{author}{\bibfnamefont{S.-J.} \bibnamefont{Paul}}, \bibnamefont{and}
  \bibinfo{author}{\bibfnamefont{R.}~\bibnamefont{Kienberger}},
  \emph{\bibinfo{title}{Photoemission chronoscopy of the iodoalkanes}},
  \bibinfo{journal}{Phys. Rev. A} \textbf{\bibinfo{volume}{112}},
  \bibinfo{pages}{L051104} (\bibinfo{year}{2025}).

\bibitem[{\citenamefont{Yao et~al.}(2026)\citenamefont{Yao, Landsman, Wang, Pi,
  and Fu}}]{Yao2026}
\bibinfo{author}{\bibfnamefont{L.-H.} \bibnamefont{Yao}},
  \bibinfo{author}{\bibfnamefont{A.~S.} \bibnamefont{Landsman}},
  \bibinfo{author}{\bibfnamefont{X.}~\bibnamefont{Wang}},
  \bibinfo{author}{\bibfnamefont{L.-W.} \bibnamefont{Pi}}, \bibnamefont{and}
  \bibinfo{author}{\bibfnamefont{Y.}~\bibnamefont{Fu}},
  \emph{\bibinfo{title}{Attosecond photoionization delays as time-domain
  photoelectron holography in polyatomic molecules}}, \bibinfo{journal}{Phys.
  Rev. Res.} \textbf{\bibinfo{volume}{8}}, \bibinfo{pages}{L032017}
  (\bibinfo{year}{2026}).

\bibitem[{\citenamefont{Toffoli et~al.}(2002)\citenamefont{Toffoli, Stener, and
  Decleva}}]{Toffoli2002}
\bibinfo{author}{\bibfnamefont{D.}~\bibnamefont{Toffoli}},
  \bibinfo{author}{\bibfnamefont{M.}~\bibnamefont{Stener}}, \bibnamefont{and}
  \bibinfo{author}{\bibfnamefont{P.}~\bibnamefont{Decleva}},
  \emph{\bibinfo{title}{Application of the relativistic time-dependent density
  functional theory to the photoionization of xenon}}, \bibinfo{journal}{J.
  Phys. B} \textbf{\bibinfo{volume}{35}}, \bibinfo{pages}{1275}
  (\bibinfo{year}{2002}).

\bibitem[{\citenamefont{Fukuzawa et~al.}(2019)\citenamefont{Fukuzawa, Yamada,
  Sakakibara, Tachibana, Ito, Takanashi, Nishiyama, Sakai, Nagaya, Saito
  et~al.}}]{Fukuzawa2019}
\bibinfo{author}{\bibfnamefont{H.}~\bibnamefont{Fukuzawa}},
  \bibinfo{author}{\bibfnamefont{S.}~\bibnamefont{Yamada}},
  \bibinfo{author}{\bibfnamefont{Y.}~\bibnamefont{Sakakibara}},
  \bibinfo{author}{\bibfnamefont{T.}~\bibnamefont{Tachibana}},
  \bibinfo{author}{\bibfnamefont{Y.}~\bibnamefont{Ito}},
  \bibinfo{author}{\bibfnamefont{T.}~\bibnamefont{Takanashi}},
  \bibinfo{author}{\bibfnamefont{T.}~\bibnamefont{Nishiyama}},
  \bibinfo{author}{\bibfnamefont{T.}~\bibnamefont{Sakai}},
  \bibinfo{author}{\bibfnamefont{K.}~\bibnamefont{Nagaya}},
  \bibinfo{author}{\bibfnamefont{N.}~\bibnamefont{Saito}},
  \bibnamefont{et~al.}, \emph{\bibinfo{title}{Probing gaseous molecular
  structure by molecular-frame photoelectron angular distributions}},
  \bibinfo{journal}{J. Chem. Phys.} \textbf{\bibinfo{volume}{151}},
  \bibinfo{pages}{104302} (\bibinfo{year}{2019}).

\bibitem[{\citenamefont{Delgado et~al.}(2025)\citenamefont{Delgado,
  Gonz{\'a}lez-Collado, Decleva, Palacios, and Mart{\'i}n}}]{Delgado2025}
\bibinfo{author}{\bibfnamefont{J.}~\bibnamefont{Delgado}},
  \bibinfo{author}{\bibfnamefont{C.~M.} \bibnamefont{Gonz{\'a}lez-Collado}},
  \bibinfo{author}{\bibfnamefont{P.}~\bibnamefont{Decleva}},
  \bibinfo{author}{\bibfnamefont{A.}~\bibnamefont{Palacios}}, \bibnamefont{and}
  \bibinfo{author}{\bibfnamefont{F.}~\bibnamefont{Mart{\'i}n}},
  \emph{\bibinfo{title}{Three-path interferences in the reconstruction of
  attosecond beatings by interference of two-photon transitions in molecules}},
  \bibinfo{journal}{Phys. Rev. A} \textbf{\bibinfo{volume}{111}},
  \bibinfo{pages}{063107} (\bibinfo{year}{2025}).



\bibitem[{\citenamefont{Pranjal et~al.}(2026)\citenamefont{Pranjal,
  Gonz{\'a}lez-V{\'a}zquez, Mart{\'i}n, and Bello}}]{Pranjal2026}
\bibinfo{author}{\bibfnamefont{P.}~\bibnamefont{Pranjal}},
  \bibinfo{author}{\bibfnamefont{J.}~\bibnamefont{Gonz{\'a}lez-V{\'a}zquez}},
  \bibinfo{author}{\bibfnamefont{F.}~\bibnamefont{Mart{\'i}n}},
  \bibnamefont{and} \bibinfo{author}{\bibfnamefont{R.~Y.} \bibnamefont{Bello}},
  \emph{\bibinfo{title}{Fully differential photoionization delays in the water
  molecule}}, \bibinfo{journal}{Phys. Rev. A}  (\bibinfo{year}{2026}),
  \bibinfo{note}{accepted 19 August 2026; DOI 10.1103/62tp-ysgk}.

\bibitem[{\citenamefont{Benda and Masin}(2024)}]{BendaMasin2024}
\bibinfo{author}{\bibfnamefont{J.}~\bibnamefont{Benda}} \bibnamefont{and}
  \bibinfo{author}{\bibfnamefont{Z.}~\bibnamefont{Masin}},
  \emph{\bibinfo{title}{Dipole-laser coupling delay in two-color rabbitt
  photoionization of polar molecules}}, \bibinfo{journal}{Phys. Rev. A}
  \textbf{\bibinfo{volume}{109}}, \bibinfo{pages}{013106}
  (\bibinfo{year}{2024}).

\bibitem[{\citenamefont{Makos et~al.}(2025)\citenamefont{Makos, Busto, Benda,
  Ertel, Merzuk, Steiner, Frassetto, Poletto, Schroter, Pfeifer
  et~al.}}]{Makos2025}
\bibinfo{author}{\bibfnamefont{I.}~\bibnamefont{Makos}},
  \bibinfo{author}{\bibfnamefont{D.}~\bibnamefont{Busto}},
  \bibinfo{author}{\bibfnamefont{J.}~\bibnamefont{Benda}},
  \bibinfo{author}{\bibfnamefont{D.}~\bibnamefont{Ertel}},
  \bibinfo{author}{\bibfnamefont{B.}~\bibnamefont{Merzuk}},
  \bibinfo{author}{\bibfnamefont{B.}~\bibnamefont{Steiner}},
  \bibinfo{author}{\bibfnamefont{F.}~\bibnamefont{Frassetto}},
  \bibinfo{author}{\bibfnamefont{L.}~\bibnamefont{Poletto}},
  \bibinfo{author}{\bibfnamefont{C.~D.} \bibnamefont{Schroter}},
  \bibinfo{author}{\bibfnamefont{T.}~\bibnamefont{Pfeifer}},
  \bibnamefont{et~al.}, \emph{\bibinfo{title}{Entanglement in photoionisation
  reveals the effect of ionic coupling in attosecond time delays}},
  \bibinfo{journal}{Nat. Commun.} \textbf{\bibinfo{volume}{16}},
  \bibinfo{pages}{8554} (\bibinfo{year}{2025}).

\bibitem[{\citenamefont{Kuchiev and Sheinerman}(1985)}]{Kuchiev1985}
\bibinfo{author}{\bibfnamefont{M.~Y.} \bibnamefont{Kuchiev}} \bibnamefont{and}
  \bibinfo{author}{\bibfnamefont{S.~A.} \bibnamefont{Sheinerman}},
  \emph{\bibinfo{title}{The post collision interaction in the inner-shell
  photoionisation of ar and xe}}, \bibinfo{journal}{Journal of Physics B:
  Atomic and Molecular Physics} \textbf{\bibinfo{volume}{18}},
  \bibinfo{pages}{L551} (\bibinfo{year}{1985}).

\bibitem[{\citenamefont{Kuchiev and Sheinerman}(1989)}]{Kuchiev1989}
\bibinfo{author}{\bibfnamefont{M.~Y.} \bibnamefont{Kuchiev}} \bibnamefont{and}
  \bibinfo{author}{\bibfnamefont{S.~A.} \bibnamefont{Sheinerman}},
  \emph{\bibinfo{title}{Post-collision interaction in atomic processes}},
  \bibinfo{journal}{Soviet Physics Uspekhi}
  \textbf{\bibinfo{volume}{32}}(\bibinfo{number}{7}), \bibinfo{pages}{569}
  (\bibinfo{year}{1989}).

\bibitem[{\citenamefont{Bauch and Bonitz}(2012)}]{BauchBonitz2012}
\bibinfo{author}{\bibfnamefont{S.}~\bibnamefont{Bauch}} \bibnamefont{and}
  \bibinfo{author}{\bibfnamefont{M.}~\bibnamefont{Bonitz}},
  \emph{\bibinfo{title}{Theoretical description of field-assisted postcollision
  interaction in {Auger} decay of atoms}}, \bibinfo{journal}{Phys. Rev. A}
  \textbf{\bibinfo{volume}{85}}, \bibinfo{pages}{053416}
  (\bibinfo{year}{2012}).

\bibitem[{\citenamefont{Decleva et~al.}(2022)\citenamefont{Decleva, Stener, and
  Toffoli}}]{Decleva2022Tiresia}
\bibinfo{author}{\bibfnamefont{P.}~\bibnamefont{Decleva}},
  \bibinfo{author}{\bibfnamefont{M.}~\bibnamefont{Stener}}, \bibnamefont{and}
  \bibinfo{author}{\bibfnamefont{D.}~\bibnamefont{Toffoli}},
  \emph{\bibinfo{title}{Continuum electronic states: The {Tiresia} code}},
  \bibinfo{journal}{Molecules}
  \textbf{\bibinfo{volume}{27}}(\bibinfo{number}{6}), \bibinfo{pages}{2026}
  (\bibinfo{year}{2022}).

\bibitem[{\citenamefont{Toffoli et~al.}(2024)\citenamefont{Toffoli, Coriani,
  Stener, and Decleva}}]{Toffoli2024Tiresia}
\bibinfo{author}{\bibfnamefont{D.}~\bibnamefont{Toffoli}},
  \bibinfo{author}{\bibfnamefont{S.}~\bibnamefont{Coriani}},
  \bibinfo{author}{\bibfnamefont{M.}~\bibnamefont{Stener}}, \bibnamefont{and}
  \bibinfo{author}{\bibfnamefont{P.}~\bibnamefont{Decleva}},
  \emph{\bibinfo{title}{{Tiresia}: A code for molecular electronic continuum
  states and photoionization}}, \bibinfo{journal}{Comput. Phys. Commun.}
  \textbf{\bibinfo{volume}{297}}, \bibinfo{pages}{109038}
  (\bibinfo{year}{2024}).

\bibitem[{\citenamefont{Stener et~al.}(2005)\citenamefont{Stener, Fronzoni, and
  Decleva}}]{jcp_1.1937367}
\bibinfo{author}{\bibfnamefont{M.}~\bibnamefont{Stener}},
  \bibinfo{author}{\bibfnamefont{G.}~\bibnamefont{Fronzoni}}, \bibnamefont{and}
  \bibinfo{author}{\bibfnamefont{P.}~\bibnamefont{Decleva}},
  \emph{\bibinfo{title}{Time-dependent density-functional theory for molecular
  photoionization with noniterative algorithm and multicenter b-spline basis
  set: Cs2 and c6h6 case studies}}, \bibinfo{journal}{The Journal of Chemical
  Physics} \textbf{\bibinfo{volume}{122}}(\bibinfo{number}{23}),
  \bibinfo{pages}{234301} (\bibinfo{year}{2005}).

\bibitem[{\citenamefont{Stener et~al.}(2006)\citenamefont{Stener, Toffoli,
  Fronzoni, and Decleva}}]{jcp_1.2178799}
\bibinfo{author}{\bibfnamefont{M.}~\bibnamefont{Stener}},
  \bibinfo{author}{\bibfnamefont{D.}~\bibnamefont{Toffoli}},
  \bibinfo{author}{\bibfnamefont{G.}~\bibnamefont{Fronzoni}}, \bibnamefont{and}
  \bibinfo{author}{\bibfnamefont{P.}~\bibnamefont{Decleva}},
  \emph{\bibinfo{title}{Time dependent density functional study of the
  photoionization dynamics of sf6}}, \bibinfo{journal}{The Journal of Chemical
  Physics} \textbf{\bibinfo{volume}{124}}(\bibinfo{number}{11}),
  \bibinfo{pages}{114306} (\bibinfo{year}{2006}).

\bibitem[{\citenamefont{te~Velde et~al.}(2001)\citenamefont{te~Velde,
  Bickelhaupt, Baerends, Fonseca~Guerra, van Gisbergen, Snijders, and
  Ziegler}}]{ADF2001}
\bibinfo{author}{\bibfnamefont{G.}~\bibnamefont{te~Velde}},
  \bibinfo{author}{\bibfnamefont{F.~M.} \bibnamefont{Bickelhaupt}},
  \bibinfo{author}{\bibfnamefont{E.~J.} \bibnamefont{Baerends}},
  \bibinfo{author}{\bibfnamefont{C.}~\bibnamefont{Fonseca~Guerra}},
  \bibinfo{author}{\bibfnamefont{S.~J.~A.} \bibnamefont{van Gisbergen}},
  \bibinfo{author}{\bibfnamefont{J.~G.} \bibnamefont{Snijders}},
  \bibnamefont{and} \bibinfo{author}{\bibfnamefont{T.}~\bibnamefont{Ziegler}},
  \emph{\bibinfo{title}{Chemistry with {ADF}}}, \bibinfo{journal}{J. Comput.
  Chem.} \textbf{\bibinfo{volume}{22}}, \bibinfo{pages}{931}
  (\bibinfo{year}{2001}).

\bibitem[{\citenamefont{{SCM}}(2022)}]{AMS2022}
\bibinfo{author}{\bibnamefont{{SCM}}}, \emph{\bibinfo{title}{{ADF} 2022.2,
  amsterdam modeling suite}}, \bibinfo{howpublished}{Theoretical Chemistry,
  Vrije Universiteit, Amsterdam, The Netherlands, \url{http://www.scm.com}}
  (\bibinfo{year}{2022}).

\bibitem[{\citenamefont{Becke}(1993)}]{Becke_1993}
\bibinfo{author}{\bibfnamefont{A.~D.} \bibnamefont{Becke}},
  \emph{\bibinfo{title}{Density‐functional thermochemistry. iii. the role of
  exact exchange}}, \bibinfo{journal}{The Journal of Chemical Physics}
  \textbf{\bibinfo{volume}{98}}(\bibinfo{number}{7}), \bibinfo{pages}{5648}
  (\bibinfo{year}{1993}).

\bibitem[{\citenamefont{Stephens et~al.}(1994)\citenamefont{Stephens, Devlin,
  Ashvar, Chabalowski, and Frisch}}]{Stephens_1994}
\bibinfo{author}{\bibfnamefont{P.~J.} \bibnamefont{Stephens}},
  \bibinfo{author}{\bibfnamefont{F.~J.} \bibnamefont{Devlin}},
  \bibinfo{author}{\bibfnamefont{C.~S.} \bibnamefont{Ashvar}},
  \bibinfo{author}{\bibfnamefont{C.~F.} \bibnamefont{Chabalowski}},
  \bibnamefont{and} \bibinfo{author}{\bibfnamefont{M.~J.}
  \bibnamefont{Frisch}}, \emph{\bibinfo{title}{Theoretical calculation of
  vibrational circular dichroism spectra}}, \bibinfo{journal}{Faraday Discuss.}
  \textbf{\bibinfo{volume}{99}}, \bibinfo{pages}{103} (\bibinfo{year}{1994}).

\bibitem[{\citenamefont{van Leeuwen and Baerends}(1994)}]{PhysRevA.49.2421}
\bibinfo{author}{\bibfnamefont{R.}~\bibnamefont{van Leeuwen}} \bibnamefont{and}
  \bibinfo{author}{\bibfnamefont{E.~J.} \bibnamefont{Baerends}},
  \emph{\bibinfo{title}{Exchange-correlation potential with correct asymptotic
  behavior}}, \bibinfo{journal}{Phys. Rev. A} \textbf{\bibinfo{volume}{49}},
  \bibinfo{pages}{2421} (\bibinfo{year}{1994}).

\bibitem[{\citenamefont{de~Boor}(1978)}]{deBoor_book}
\bibinfo{author}{\bibfnamefont{C.}~\bibnamefont{de~Boor}},
  \emph{\bibinfo{title}{A Practical Guide to Splines}}
  (\bibinfo{publisher}{Springer}, \bibinfo{address}{New York},
  \bibinfo{year}{1978}).

\bibitem[{\citenamefont{Brosolo and Decleva}(1992)}]{BROSOLO1992185}
\bibinfo{author}{\bibfnamefont{M.}~\bibnamefont{Brosolo}} \bibnamefont{and}
  \bibinfo{author}{\bibfnamefont{P.}~\bibnamefont{Decleva}},
  \emph{\bibinfo{title}{Variational approach to continuum orbitals in a spline
  basis: An application to h2+ photoionization}}, \bibinfo{journal}{Chemical
  Physics} \textbf{\bibinfo{volume}{159}}(\bibinfo{number}{2}),
  \bibinfo{pages}{185} (\bibinfo{year}{1992}).

\bibitem[{\citenamefont{Zangwill and Soven}(1980)}]{PhysRevA.21.1561}
\bibinfo{author}{\bibfnamefont{A.}~\bibnamefont{Zangwill}} \bibnamefont{and}
  \bibinfo{author}{\bibfnamefont{P.}~\bibnamefont{Soven}},
  \emph{\bibinfo{title}{Density-functional approach to local-field effects in
  finite systems: Photoabsorption in the rare gases}}, \bibinfo{journal}{Phys.
  Rev. A} \textbf{\bibinfo{volume}{21}}, \bibinfo{pages}{1561}
  (\bibinfo{year}{1980}).

\bibitem[{\citenamefont{Gross and Kohn}(1990)}]{GROSS1990255}
\bibinfo{author}{\bibfnamefont{E.}~\bibnamefont{Gross}} \bibnamefont{and}
  \bibinfo{author}{\bibfnamefont{W.}~\bibnamefont{Kohn}}, in
  \emph{\bibinfo{booktitle}{Density Functional Theory of Many-Fermion
  Systems}}, edited by \bibinfo{editor}{\bibfnamefont{P.-O.}
  \bibnamefont{L\"owdin}} (\bibinfo{publisher}{Academic Press},
  \bibinfo{year}{1990}), vol.~\bibinfo{volume}{21} of
  \emph{\bibinfo{series}{Advances in Quantum Chemistry}}, pp.
  \bibinfo{pages}{255--291},
  \urlprefix\url{https://www.sciencedirect.com/science/article/pii/S0065327608606000}.

\bibitem[{\citenamefont{Chacon et~al.}(2014)\citenamefont{Chacon, Lein, and
  Ruiz}}]{Chacon2014}
\bibinfo{author}{\bibfnamefont{A.}~\bibnamefont{Chacon}},
  \bibinfo{author}{\bibfnamefont{M.}~\bibnamefont{Lein}}, \bibnamefont{and}
  \bibinfo{author}{\bibfnamefont{C.}~\bibnamefont{Ruiz}},
  \emph{\bibinfo{title}{Asymmetry of wigner's time delay in a small molecule}},
  \bibinfo{journal}{Phys. Rev. A} \textbf{\bibinfo{volume}{89}},
  \bibinfo{pages}{053427} (\bibinfo{year}{2014}).

\bibitem[{\citenamefont{Hockett et~al.}(2016)\citenamefont{Hockett, Frumker,
  Villeneuve, and Corkum}}]{Hockett2016}
\bibinfo{author}{\bibfnamefont{P.}~\bibnamefont{Hockett}},
  \bibinfo{author}{\bibfnamefont{E.}~\bibnamefont{Frumker}},
  \bibinfo{author}{\bibfnamefont{D.~M.} \bibnamefont{Villeneuve}},
  \bibnamefont{and} \bibinfo{author}{\bibfnamefont{P.~B.}
  \bibnamefont{Corkum}}, \emph{\bibinfo{title}{Time delay in molecular
  photoionization}}, \bibinfo{journal}{J. Phys. B: At. Mol. Opt. Phys.}
  \textbf{\bibinfo{volume}{49}}, \bibinfo{pages}{095602}
  (\bibinfo{year}{2016}).

\bibitem[{\citenamefont{Baykusheva and W{\"o}rner}(2017)}]{Baykusheva2017}
\bibinfo{author}{\bibfnamefont{D.}~\bibnamefont{Baykusheva}} \bibnamefont{and}
  \bibinfo{author}{\bibfnamefont{H.~J.} \bibnamefont{W{\"o}rner}},
  \emph{\bibinfo{title}{Theory of attosecond delays in molecular
  photoionization}}, \bibinfo{journal}{J. Chem. Phys.}
  \textbf{\bibinfo{volume}{146}}, \bibinfo{pages}{124306}
  (\bibinfo{year}{2017}).

\bibitem[{\citenamefont{Serov and Kheifets}(2017)}]{Serov2017}
\bibinfo{author}{\bibfnamefont{V.~V.} \bibnamefont{Serov}} \bibnamefont{and}
  \bibinfo{author}{\bibfnamefont{A.~S.} \bibnamefont{Kheifets}},
  \emph{\bibinfo{title}{Time delay in {XUV/IR} photoionization of {H$_2$O}}},
  \bibinfo{journal}{J. Chem. Phys.} \textbf{\bibinfo{volume}{147}},
  \bibinfo{pages}{204303} (\bibinfo{year}{2017}).

\bibitem[{\citenamefont{Borr{\`a}s et~al.}(2023)\citenamefont{Borr{\`a}s,
  Gonz{\'a}lez-V{\'a}zquez, Argenti, and Mart{\'i}n}}]{Borras2023}
\bibinfo{author}{\bibfnamefont{V.~J.} \bibnamefont{Borr{\`a}s}},
  \bibinfo{author}{\bibfnamefont{J.}~\bibnamefont{Gonz{\'a}lez-V{\'a}zquez}},
  \bibinfo{author}{\bibfnamefont{L.}~\bibnamefont{Argenti}}, \bibnamefont{and}
  \bibinfo{author}{\bibfnamefont{F.}~\bibnamefont{Mart{\'i}n}},
  \emph{\bibinfo{title}{Attosecond photoionization delays in the vicinity of
  molecular feshbach resonances}}, \bibinfo{journal}{Sci. Adv.}
  \textbf{\bibinfo{volume}{9}}, \bibinfo{pages}{eade3855}
  (\bibinfo{year}{2023}).

\bibitem[{\citenamefont{Su{\~n}er-Rubio
  et~al.}(2024)\citenamefont{Su{\~n}er-Rubio, Lemell, Bello, Burgd{\"o}rfer,
  Palacios, and Mart{\'i}n}}]{SunerRubio2024}
\bibinfo{author}{\bibfnamefont{A.~J.} \bibnamefont{Su{\~n}er-Rubio}},
  \bibinfo{author}{\bibfnamefont{C.}~\bibnamefont{Lemell}},
  \bibinfo{author}{\bibfnamefont{R.~Y.} \bibnamefont{Bello}},
  \bibinfo{author}{\bibfnamefont{J.}~\bibnamefont{Burgd{\"o}rfer}},
  \bibinfo{author}{\bibfnamefont{A.}~\bibnamefont{Palacios}}, \bibnamefont{and}
  \bibinfo{author}{\bibfnamefont{F.}~\bibnamefont{Mart{\'i}n}},
  \emph{\bibinfo{title}{Attosecond photoionization delays in molecules: The
  role of nuclear motion}}, \bibinfo{journal}{Phys. Rev. Research}
  \textbf{\bibinfo{volume}{6}}, \bibinfo{pages}{L022066}
  (\bibinfo{year}{2024}).

\bibitem[{\citenamefont{Gonz{\'a}lez-Collado
  et~al.}(2024)\citenamefont{Gonz{\'a}lez-Collado, Cattaneo, Pl{\'e}siat,
  Omiste, Vos, Gonz{\'a}lez-V{\'a}zquez, Decleva, Keller, Palacios, and
  Mart{\'{\i}}n}}]{GonzalezCollado2024}
\bibinfo{author}{\bibfnamefont{C.~M.} \bibnamefont{Gonz{\'a}lez-Collado}},
  \bibinfo{author}{\bibfnamefont{L.}~\bibnamefont{Cattaneo}},
  \bibinfo{author}{\bibfnamefont{E.}~\bibnamefont{Pl{\'e}siat}},
  \bibinfo{author}{\bibfnamefont{J.~J.} \bibnamefont{Omiste}},
  \bibinfo{author}{\bibfnamefont{J.}~\bibnamefont{Vos}},
  \bibinfo{author}{\bibfnamefont{J.}~\bibnamefont{Gonz{\'a}lez-V{\'a}zquez}},
  \bibinfo{author}{\bibfnamefont{P.}~\bibnamefont{Decleva}},
  \bibinfo{author}{\bibfnamefont{U.}~\bibnamefont{Keller}},
  \bibinfo{author}{\bibfnamefont{A.}~\bibnamefont{Palacios}}, \bibnamefont{and}
  \bibinfo{author}{\bibfnamefont{F.}~\bibnamefont{Mart{\'{\i}}n}},
  \emph{\bibinfo{title}{Orientation-resolved attosecond photoionization delays
  in the {N$_2$O} molecule}}, \bibinfo{journal}{Phys. Rev. Research}
  \textbf{\bibinfo{volume}{6}}, \bibinfo{pages}{033342} (\bibinfo{year}{2024}).

\bibitem[{\citenamefont{Rescigno et~al.}(2024)\citenamefont{Rescigno,
      Trevisan,
  Lucchese, and McCurdy}}]{Rescigno2024}
\bibinfo{author}{\bibfnamefont{T.~N.} \bibnamefont{Rescigno}},
  \bibinfo{author}{\bibfnamefont{C.~S.} \bibnamefont{Trevisan}},
  \bibinfo{author}{\bibfnamefont{R.~R.} \bibnamefont{Lucchese}},
  \bibnamefont{and} \bibinfo{author}{\bibfnamefont{C.~W.}
  \bibnamefont{McCurdy}}, \emph{\bibinfo{title}{Attosecond time delays
      in the
  body frame for photodetachment from molecular anions}},
  \bibinfo{journal}{Phys. Rev. A} \textbf{\bibinfo{volume}{110}},
  \bibinfo{pages}{013122} (\bibinfo{year}{2024}).

\bibitem[{\citenamefont{Pranjal et~al.}(2025)\citenamefont{Pranjal,
  Gonz{\'a}lez-V{\'a}zquez, Bello, and Mart{\'i}n}}]{Pranjal2025}
\bibinfo{author}{\bibfnamefont{P.}~\bibnamefont{Pranjal}},
  \bibinfo{author}{\bibfnamefont{J.}~\bibnamefont{Gonz{\'a}lez-V{\'a}zquez}},
  \bibinfo{author}{\bibfnamefont{R.~Y.} \bibnamefont{Bello}}, \bibnamefont{and}
  \bibinfo{author}{\bibfnamefont{F.}~\bibnamefont{Mart{\'i}n}},
  \emph{\bibinfo{title}{Vibrationally resolved photoionization delays in the
  water molecule}}, \bibinfo{journal}{Phys. Rev. Lett.}
  \textbf{\bibinfo{volume}{135}}, \bibinfo{pages}{223202}
  (\bibinfo{year}{2025}).

\bibitem[{\citenamefont{Kheifets}(2013{\natexlab{a}})}]{Kheifets2013}
\bibinfo{author}{\bibfnamefont{A.~S.} \bibnamefont{Kheifets}},
  \emph{\bibinfo{title}{Time delay in valence-shell photoionization of
  noble-gas atoms}}, \bibinfo{journal}{Phys. Rev. A}
  \textbf{\bibinfo{volume}{87}}, \bibinfo{pages}{063404}
  (\bibinfo{year}{2013}{\natexlab{a}}).

\bibitem[{\citenamefont{Kheifets et~al.}(2020)\citenamefont{Kheifets, Toffoli,
  and Decleva}}]{Kheifets2020}
\bibinfo{author}{\bibfnamefont{A.~S.} \bibnamefont{Kheifets}},
  \bibinfo{author}{\bibfnamefont{D.}~\bibnamefont{Toffoli}}, \bibnamefont{and}
  \bibinfo{author}{\bibfnamefont{P.}~\bibnamefont{Decleva}},
  \emph{\bibinfo{title}{Angular dependent time delay near correlation induced
  cooper minima}}, \bibinfo{journal}{J. Phys. B} \textbf{\bibinfo{volume}{53}},
  \bibinfo{pages}{115201} (\bibinfo{year}{2020}).

\bibitem[{\citenamefont{Grafstrom and Landsman}(2023)}]{Grafstrom2023}
\bibinfo{author}{\bibfnamefont{B.}~\bibnamefont{Grafstrom}} \bibnamefont{and}
  \bibinfo{author}{\bibfnamefont{A.~S.} \bibnamefont{Landsman}},
  \emph{\bibinfo{title}{Attosecond time delay trends across the isoelectronic
  noble gas sequence}}, \bibinfo{journal}{Atoms} \textbf{\bibinfo{volume}{11}},
  \bibinfo{pages}{84} (\bibinfo{year}{2023}).

\bibitem[{\citenamefont{Banerjee et~al.}(2021)\citenamefont{Banerjee, Aarthi,
  Saha, Aravind, and Deshmukh}}]{Banerjee2021}
\bibinfo{author}{\bibfnamefont{S.}~\bibnamefont{Banerjee}},
  \bibinfo{author}{\bibfnamefont{G.}~\bibnamefont{Aarthi}},
  \bibinfo{author}{\bibfnamefont{S.}~\bibnamefont{Saha}},
  \bibinfo{author}{\bibfnamefont{G.}~\bibnamefont{Aravind}}, \bibnamefont{and}
  \bibinfo{author}{\bibfnamefont{P.~C.} \bibnamefont{Deshmukh}},
  \emph{\bibinfo{title}{Time delay in negative ion photodetachment}},
  \bibinfo{journal}{Phys. Scr.} \textbf{\bibinfo{volume}{96}},
  \bibinfo{pages}{114005} (\bibinfo{year}{2021}).

\bibitem[{\citenamefont{Forster et~al.}(2026)\citenamefont{Forster, Pollanka,
  Paul, Freisinger, Schr{\"o}der, Richter, Alagia, Bruno, Stranges, and
  Kienberger}}]{Forster2026}
\bibinfo{author}{\bibfnamefont{M.}~\bibnamefont{Forster}},
  \bibinfo{author}{\bibfnamefont{M.}~\bibnamefont{Pollanka}},
  \bibinfo{author}{\bibfnamefont{S.-J.} \bibnamefont{Paul}},
  \bibinfo{author}{\bibfnamefont{P.}~\bibnamefont{Freisinger}},
  \bibinfo{author}{\bibfnamefont{C.}~\bibnamefont{Schr{\"o}der}},
  \bibinfo{author}{\bibfnamefont{R.}~\bibnamefont{Richter}},
  \bibinfo{author}{\bibfnamefont{M.}~\bibnamefont{Alagia}},
  \bibinfo{author}{\bibfnamefont{A.}~\bibnamefont{Bruno}},
  \bibinfo{author}{\bibfnamefont{S.}~\bibnamefont{Stranges}}, \bibnamefont{and}
  \bibinfo{author}{\bibfnamefont{R.}~\bibnamefont{Kienberger}},
  \emph{\bibinfo{title}{Photoemission study of iodoalkanes}},
  \bibinfo{journal}{Phys. Rev. A}  (\bibinfo{year}{2026}), \bibinfo{note}{to be
  submitted}.

\bibitem[{\citenamefont{O'Sullivan}(1982)}]{OSullivan1982}
\bibinfo{author}{\bibfnamefont{G.}~\bibnamefont{O'Sullivan}},
  \emph{\bibinfo{title}{The absorption spectrum of CH$_3$I in the extreme vuv}},
  \bibinfo{journal}{Journal of Physics B: Atomic and Molecular Physics}
  \textbf{\bibinfo{volume}{15}}, \bibinfo{pages}{L327} (\bibinfo{year}{1982}).

\bibitem[{\citenamefont{Lindle et~al.}(1984)\citenamefont{Lindle, Kobrin,
  Truesdale, Ferrett, Heimann, Kerkhoff, Becker, and Shirley}}]{Lindle1984}
\bibinfo{author}{\bibfnamefont{D.~W.} \bibnamefont{Lindle}},
  \bibinfo{author}{\bibfnamefont{P.~H.} \bibnamefont{Kobrin}},
  \bibinfo{author}{\bibfnamefont{C.~M.} \bibnamefont{Truesdale}},
  \bibinfo{author}{\bibfnamefont{T.~A.} \bibnamefont{Ferrett}},
  \bibinfo{author}{\bibfnamefont{P.~A.} \bibnamefont{Heimann}},
  \bibinfo{author}{\bibfnamefont{H.~G.} \bibnamefont{Kerkhoff}},
  \bibinfo{author}{\bibfnamefont{U.}~\bibnamefont{Becker}}, \bibnamefont{and}
  \bibinfo{author}{\bibfnamefont{D.~A.} \bibnamefont{Shirley}},
  \emph{\bibinfo{title}{Inner-shell photoemission from the iodine atom in
  {CH$_3$I}}}, \bibinfo{journal}{Phys. Rev. A} \textbf{\bibinfo{volume}{30}},
  \bibinfo{pages}{239} (\bibinfo{year}{1984}).

\bibitem[{\citenamefont{Kheifets and Catsamas}(2023)}]{PhysRevA.107.L021102}
\bibinfo{author}{\bibfnamefont{A.~S.} \bibnamefont{Kheifets}} \bibnamefont{and}
  \bibinfo{author}{\bibfnamefont{S.}~\bibnamefont{Catsamas}},
  \emph{\bibinfo{title}{Shape resonances in photoionization cross sections and
  time delay}}, \bibinfo{journal}{Phys. Rev. A} \textbf{\bibinfo{volume}{107}},
  \bibinfo{pages}{L021102} (\bibinfo{year}{2023}).

\bibitem[{\citenamefont{Ji et~al.}(2024)\citenamefont{Ji, Kheifets, Han, Ueda,
  and W\"orner}}]{Ji2024}
\bibinfo{author}{\bibfnamefont{J.-B.} \bibnamefont{Ji}},
  \bibinfo{author}{\bibfnamefont{A.~S.} \bibnamefont{Kheifets}},
  \bibinfo{author}{\bibfnamefont{M.}~\bibnamefont{Han}},
  \bibinfo{author}{\bibfnamefont{K.}~\bibnamefont{Ueda}}, \bibnamefont{and}
  \bibinfo{author}{\bibfnamefont{H.~J.} \bibnamefont{W\"orner}},
  \emph{\bibinfo{title}{Relation between photoionisation cross sections and
  attosecond time delays}}, \bibinfo{journal}{New J. Phys.}
  \textbf{\bibinfo{volume}{26}}(\bibinfo{number}{9}), \bibinfo{pages}{093014}
  (\bibinfo{year}{2024}).

\bibitem[{\citenamefont{Kheifets}(2025)}]{Kheifets2025resonant}
\bibinfo{author}{\bibfnamefont{A.~S.} \bibnamefont{Kheifets}},
  \emph{\bibinfo{title}{Resonant photoionization and time delay. {Topical}
  review}}, \bibinfo{journal}{J.~Phys.~B}
  \textbf{\bibinfo{volume}{58}}(\bibinfo{number}{7}), \bibinfo{pages}{072001}
  (\bibinfo{year}{2025}).

\bibitem[{\citenamefont{Driver et~al.}(2024)\citenamefont{Driver, Mountney,
  Wang, Ortmann, Al-Haddad, Berrah, Bostedt, Champenois, DiMauro, Duris
  et~al.}}]{Driver2024}
\bibinfo{author}{\bibfnamefont{T.}~\bibnamefont{Driver}},
  \bibinfo{author}{\bibfnamefont{M.}~\bibnamefont{Mountney}},
  \bibinfo{author}{\bibfnamefont{J.}~\bibnamefont{Wang}},
  \bibinfo{author}{\bibfnamefont{L.}~\bibnamefont{Ortmann}},
  \bibinfo{author}{\bibfnamefont{A.}~\bibnamefont{Al-Haddad}},
  \bibinfo{author}{\bibfnamefont{N.}~\bibnamefont{Berrah}},
  \bibinfo{author}{\bibfnamefont{C.}~\bibnamefont{Bostedt}},
  \bibinfo{author}{\bibfnamefont{E.~G.} \bibnamefont{Champenois}},
  \bibinfo{author}{\bibfnamefont{L.~F.} \bibnamefont{DiMauro}},
  \bibinfo{author}{\bibfnamefont{J.}~\bibnamefont{Duris}},
  \bibnamefont{et~al.}, \emph{\bibinfo{title}{Attosecond delays in x-ray
  molecular ionization}}, \bibinfo{journal}{Nature}
  \textbf{\bibinfo{volume}{632}}(\bibinfo{number}{8026}), \bibinfo{pages}{762}
  (\bibinfo{year}{2024}).


\bibitem[{\citenamefont{Kuchiev and Sheinerman}(1986)}]{Kuchiev1986CPC}
\bibinfo{author}{\bibfnamefont{M.~Y.} \bibnamefont{Kuchiev}} \bibnamefont{and}
  \bibinfo{author}{\bibfnamefont{S.~A.} \bibnamefont{Sheinerman}},
  \emph{\bibinfo{title}{The calculation of the photoionization cross section
  allowing for the post collision interaction}}, \bibinfo{journal}{Computer
  Physics Communications} \textbf{\bibinfo{volume}{39}}, \bibinfo{pages}{155}
  (\bibinfo{year}{1986}).

\bibitem[{\citenamefont{Kuchiev and Sheinerman}(1988)}]{Kuchiev1988}
\bibinfo{author}{\bibfnamefont{M.~Y.} \bibnamefont{Kuchiev}} \bibnamefont{and}
  \bibinfo{author}{\bibfnamefont{S.~A.} \bibnamefont{Sheinerman}},
  \emph{\bibinfo{title}{Resonant scattering with low-velocity outgoing charged
  particles}}, \bibinfo{journal}{Journal of Physics B: Atomic, Molecular and
  Optical Physics} \textbf{\bibinfo{volume}{21}}, \bibinfo{pages}{2027}
  (\bibinfo{year}{1988}).

\bibitem[{\citenamefont{Hasan et~al.}(2026)\citenamefont{Hasan, Gao, Liang,
  Yuan, Eisenhutt, Tsai, Chen, W{\"o}rner, Rudenko, and Han}}]{Hasan2026}
\bibinfo{author}{\bibfnamefont{M.}~\bibnamefont{Hasan}},
  \bibinfo{author}{\bibfnamefont{J.}~\bibnamefont{Gao}},
  \bibinfo{author}{\bibfnamefont{H.}~\bibnamefont{Liang}},
  \bibinfo{author}{\bibfnamefont{Y.}~\bibnamefont{Yuan}},
  \bibinfo{author}{\bibfnamefont{Z.}~\bibnamefont{Eisenhutt}},
  \bibinfo{author}{\bibfnamefont{M.-S.} \bibnamefont{Tsai}},
  \bibinfo{author}{\bibfnamefont{M.-C.} \bibnamefont{Chen}},
  \bibinfo{author}{\bibfnamefont{H.~J.} \bibnamefont{W{\"o}rner}},
  \bibinfo{author}{\bibfnamefont{A.}~\bibnamefont{Rudenko}}, \bibnamefont{and}
  \bibinfo{author}{\bibfnamefont{M.}~\bibnamefont{Han}},
  \emph{\bibinfo{title}{Time-domain measurement of auger electron dynamics in
  xenon and krypton atoms after giant resonance photoionization}},
  \bibinfo{journal}{Physical Review Letters} \textbf{\bibinfo{volume}{136}},
  \bibinfo{pages}{163201} (\bibinfo{year}{2026}).

\bibitem[{\citenamefont{Carneiro et~al.}(2026)\citenamefont{Carneiro, Lepetit,
  Crippa, Ciuffarella, Marroux, Hergott, Palaudoux, Gu{\'e}naux, Tcherbakoff,
  Sali{\`e}res et~al.}}]{Carneiro2026}
\bibinfo{author}{\bibfnamefont{G.}~\bibnamefont{Carneiro}},
  \bibinfo{author}{\bibfnamefont{F.}~\bibnamefont{Lepetit}},
  \bibinfo{author}{\bibfnamefont{G.}~\bibnamefont{Crippa}},
  \bibinfo{author}{\bibfnamefont{G.}~\bibnamefont{Ciuffarella}},
  \bibinfo{author}{\bibfnamefont{H.~J.~B.} \bibnamefont{Marroux}},
  \bibinfo{author}{\bibfnamefont{J.-F.} \bibnamefont{Hergott}},
  \bibinfo{author}{\bibfnamefont{J.}~\bibnamefont{Palaudoux}},
  \bibinfo{author}{\bibfnamefont{N.}~\bibnamefont{Gu{\'e}naux}},
  \bibinfo{author}{\bibfnamefont{O.}~\bibnamefont{Tcherbakoff}},
  \bibinfo{author}{\bibfnamefont{P.}~\bibnamefont{Sali{\`e}res}},
  \bibnamefont{et~al.}, \emph{\bibinfo{title}{Attosecond spectroscopy of
  iodomethane}}, in \emph{\bibinfo{booktitle}{ELIXIR 2026 Book of Abstracts}}
  (\bibinfo{address}{Synchrotron SOLEIL, France}, \bibinfo{year}{2026}),
  \bibinfo{note}{poster PO-02, ELIXIR 2026, March 9--11, 2026}.

\bibitem[{\citenamefont{Kheifets et~al.}(2026)\citenamefont{Kheifets, Toffoli,
  and Decleva}}]{Kheifets2026Zenodo}
\bibinfo{author}{\bibfnamefont{A.~S.} \bibnamefont{Kheifets}},
  \bibinfo{author}{\bibfnamefont{D.}~\bibnamefont{Toffoli}}, \bibnamefont{and}
  \bibinfo{author}{\bibfnamefont{P.}~\bibnamefont{Decleva}},
  \emph{\bibinfo{title}{{TDDFT} photoionization cross sections, anisotropy
  parameters, and {Wigner} time delays for iodoalkanes}}
  (\bibinfo{year}{2026}),
  \urlprefix\url{https://doi.org/10.5281/zenodo.23092656}.

\end{thebibliography}

\end{document}